\documentclass[preprint]{aastex}
\usepackage{amsmath, amsfonts,amssymb,natbib,graphicx,subfig,rotating,lscape,multirow,color}

\begin{document}

\title{Only the Lonely: H \textsc{I} Imaging of Void Galaxies}
\shorttitle{H \textsc{I} Imaging of Void Galaxies}

\author{K. Kreckel\altaffilmark{1}, E. Platen\altaffilmark{2}, M. A. Arag\'on-Calvo\altaffilmark{3}, J. H. van Gorkom\altaffilmark{1}, R. van de Weygaert\altaffilmark{2}, J. M. van der Hulst\altaffilmark{2},  K. Kova\v{c}\altaffilmark{4}, C.-W. Yip\altaffilmark{3}, P. J. E. Peebles\altaffilmark{5}}

\altaffiltext{1}{Department of Astronomy, Columbia University, Mail Code 5246, 550 West 120th Street, New York, NY 10027, USA; email: kstanonik@astro.columbia.edu}
\altaffiltext{2}{Kapteyn Astronomical Institute, University of Groningen, PO Box 800, 9700 AV Groningen, the Netherlands}
\altaffiltext{3}{The Johns Hopkins University, 3701 San Martin Drive, Baltimore, MD 21218, USA}
\altaffiltext{4}{Institute of Astronomy, Swiss Federal Institute of Technology (ETH H\"{o}nggerberg), CH-8093, Z\"{u}rich, Switzerland}
\altaffiltext{5}{Joseph Henry Laboratories, Princeton University, Princeton, NJ 08544, USA}


\begin{abstract}
Void galaxies, residing within the deepest underdensities of the Cosmic Web, present an ideal population for the study of 
galaxy formation and evolution in an environment undisturbed by the complex processes modifying galaxies in clusters and groups, as well as provide an observational test for theories of cosmological structure formation.  We have completed a pilot survey for the H \textsc{i} imaging aspects of a new Void Galaxy Survey (VGS), imaging 15 void galaxies in H \textsc{i} in local ($d < 100$ Mpc) voids.  H \textsc{i} masses range from $3.5 \times 10^8$ to $3.8 \times 10^9 M_\odot$, with one nondetection with an upper limit of  $2.1 \times 10^8 M_\odot$.  Our galaxies were selected using a structural and geometric technique to produce a sample that is purely environmentally selected and uniformly represents the void galaxy population.  In addition, we use a powerful new backend of the Westerbork Synthesis Radio Telescope that allows us to probe a large volume around each targeted galaxy, simultaneously providing an environmentally constrained sample of fore- and background control sample of galaxies while still resolving individual galaxy kinematics and detecting faint companions in H \textsc{i}.  This small sample makes up a surprisingly interesting collection of perturbed and interacting galaxies, all with small stellar disks.  Four galaxies have significantly perturbed H \textsc{i} disks, five have previously unidentified companions at distances ranging from 50 to 200 kpc, two are in interacting systems, and one was found to have a polar H \textsc{i} disk.  
Our initial findings suggest void galaxies are a gas-rich, dynamic population which present evidence of ongoing gas accretion, major and minor interactions, and filamentary alignment despite the surrounding underdense environment.
\end{abstract}

\section{Introduction} 
\label{sec:intro}
With the prevalence of ever wider and deeper redshift surveys, from the second Center for Astrophysics Redshift Survey \citep{Huchra1983} to 
the Sloan Digital Sky Survey (SDSS, \citealt{York2000}), we have refined our ability to identify the elongated 
filaments, sheetlike walls and dense compact clusters that compose the Cosmic Web \citep{Zeldovich1970,
Klypin1983,Lapparent1986,Bond1996,weybond2009}.  These structures surround voids, enormous regions 
$10-30h^{-1}$ Mpc in diameter that are largely devoid of galaxies and occupy most 
of the volume of the Universe (\citealt{Gregory1978,Einasto1980,Kirshner1981,Lapparent1986,Hoyle2004}; for a recent review see \citealt{weyplaten2009}).
Within these underdense -- but not empty -- regions, large galaxy redshift surveys have allowed us to 
distinguish an environmentally defined population of void galaxies residing in regions up to 10 times less dense than 
the cosmic mean.  Largely unaffected by the complexities and processes modifying galaxies in high-density environments, 
these isolated void regions are expected to hold important clues to understanding the environmental 
influences on galaxy formation and evolution 
\citep{Szomoru1996,Grogin2000,Corbin2005,Patiri2006b,Pustilnik2006,Wegner2008,Stanonik2009}. 
Additionally, the apparent underabundance of galaxies in the void regions may present a challenge for currently 
favored galaxy formation theories 
\citep{Peebles2001,Mathis2002,Gottloeber2003,Furlanetto2006,Tinker2009},
while their distribution is expected to trace substructure within voids, tenuous features which are fossil remnants of the 
hierarchical buildup of the Cosmic Web 
\citep{Dubinski1993,Weygaert1993,Popescu1997,Shethwey2004,Patiri2006a,Tikhonov2006}.

Void galaxies appear to have a more youthful state of star formation. As a population, void galaxies are statistically bluer, have a later morphological type, and have higher 
specific star formation rates than galaxies in average density environments \citep{Grogin1999,Grogin2000,Rojas2004,Rojas2005}. 
Whether void galaxies are intrinsically different or whether their characteristics are simply 
due to the low mass bias of the galaxy luminosity function in low density regions is still an issue of discussion. 
Overall, the mean colors of the red and blue void galaxy populations, taken separately, are  comparable to galaxies 
in average density environments at the same luminosity, though an excess of blue galaxies is apparent \citep{Balogh2004,Patiri2006b}.
This suggests that in some respects the general underdensity of the environment has had little to no impact on their development, 
raising the question of to what extent the global ($\sim$20 Mpc) as opposed to local ($\sim$1 Mpc) environment shapes the 
formation and evolution of galaxies. 
\cite{Park2007} suggest that it is only the morphology and luminosity of 
void galaxies which is dependent on their environment, with all other statistical correlations stemming from these two 
key parameters. Interestingly, they and others find contradictory indications of a slight blueward shift of the 
blue cloud in voids at fixed luminosity \citep{Blanton2005,Bendabeck2008}.
No such shift has been found for the red 
sequence of early type galaxies. 

Less is known about the gas content of void galaxies.  \cite{Szomoru1996} surveyed galaxies within the  Bo\"{o}tes void with pointed observations of 24 \textit{IRAS} selected galaxies, of which 16 were detected.  Most of these galaxies
 are found to be gas-rich and disk-like, with many gas-rich companions, however more complete redshift surveys show that many of the targeted galaxies  reside in the outer realms of the void and with some reason might be identified with the moderate density environment of walls \citep[][also see Section~\ref{sec:analysis}]{Platen2009}. 
\cite{Huchtmeier1997} find that dwarf galaxies in voids have a higher $M_{\rm H~\textsc{i}}/L_B$ ratio the deeper within the underdensity they reside.
Because of the active star-forming nature of void galaxies, detailed H \textsc{i} observations are key to understanding the environmental differences observed.

The unique nature of void galaxies provides an ideal chance to distinguish the role of environment 
in gas accretion and galaxy evolution on an individual basis. 
Fresh gas accretion is necessary for galaxies to maintain star formation rates seen today without depleting their observed gas mass 
in less than a Hubble time \citep{Larson1972}. Historically, this gas was assumed to condense out of reservoirs of hot gas existing in halos 
around galaxies \citep{Rees1977,Silk1977,White1978,White1991}, with some amount 
of gas recycling  via galactic fountains \citep{Fraternali2008}.  However, recent simulations have renewed interest in 
the slow accretion of cold gas along filaments \citep{Binney1977,Keres2005,Gao2005,Dekel2006,Dekel2009}. 
Void galaxies provide a 
unique sample of younger, star forming galaxies where their inherent isolation may allow us to distinguish the effects of close 
encounters and galaxy mergers from other mechanisms of gas accretion, 
and their constrained environment allows a search for systematic trends in neutral gas content and distribution with cosmological density.  

Despite the remarkable success of $\Lambda$CDM cosmology in explaining the general cosmic matter distribution there are a few telling discrepancies, the most prominent of which concerns the over-prediction of low-mass halos within voids \citep{Mathis2002,Gottloeber2003,Furlanetto2006,Peebles2010}. The observed density  of faint ($-18<M_B<-12$) galaxies in voids is only $1/100$th  that of the mean \citep{Kuhn1997,Karachentsev2004}, in contrast to the predictions of high-resolution $\Lambda$CDM simulations that the density of low mass halos ($10^9 \hbox{\rm M}_{\odot}< \hbox{\rm M}<10^{11} \hbox{\rm M}_{\odot}$) should be $1/10$th that of the cosmic mean \citep{Warren2006,Hoeft2006}. In addition, simulations predict that these low mass halos will trail into the voids, while in deep optical surveys we see that dwarf galaxies avoid the empty regions defined by the more luminous galaxies \citep{Kuhn1997}. This phenomenon has also been noted in blind H \textsc{i} surveys \citep{Saintonge2008} and deep surveys of the local volume \citep{Tikhonov2009}, where in general most void galaxies are found at the edges of voids. \cite{Peebles2001} has strongly emphasized that this dearth of dwarf and/or low surface brightness galaxies in voids cannot be straightforwardly understood in our standard view of galaxy formation. 

There have been many solutions proposed which aim to limit galaxy formation within the least massive halos \citep{Furlanetto2006,Hoeft2006}. \cite{Tinker2009} suggest that the void phenomenon might be understood if the properties of galaxies are solely dependent on the mass of the dark halos in which they live, independent of their environment, assuming a sufficiently tailored halo occupation distribution. However, it remains difficult to explain how the implied severe degrees of bias between high- and low-luminosity galaxies at the void boundaries can be reconciled with the observations of nearby voids: there are no indications for the predicted segregation of fainter galaxies being found further into the void interior (Vogeley, private communication). It is also revealing that predictions for the galaxy distribution in void regions by different semi-analytical galaxy formation schemes, in the context of the Millennium simulation or other simulations \citep{Mathis2002,Lucia2006,Bower2006}, disagree with each other at a remarkably fundamental level \citep[see e.g.][]{Platen2009}. Ultimately, we will need a better understanding of the various gas, radiation and feedback processes, such as investigated by \cite{Hoeft2006}.  The issue remains far from solved, and progress will largely depend on new observations.

Within the context of hierarchical cosmological structure formation scenarios, voids are expected to exhibit a rich dark 
matter substructure which is a remnant of the hierarchical buildup of  voids and provides 
additional  constraints on theories of cosmological evolution \citep{Regoes1991,Weygaert1993,Shethwey2004,Colberg2005,Ceccar2006,2010MNRAS.404L..89A}. 
The merging of expanding voids dilutes the intervening substructure to cause a cosmic flow away from the void centers and 
along walls and filaments \citep{Dubinski1993,Shethwey2004}. The extent to which the tenuous void 
substructure may also be recognized in the galaxy distribution is not yet fully settled, however there may be
a link between the star formation activity of void galaxies and the influx of (coldly) 
accreting matter transported along dark matter filaments \citep{Zitrin2009,Parklee2009a,Parklee2009b}. 
Some observational studies claim to
 recognize patterns in the void galaxy distribution \citep{Popescu1997,Szomoru1996,Platen2009}. 
Moreover, there are indications that the clustering of galaxies in voids  appears to be of comparable strength to that of galaxies 
in average density environments \citep{Szomoru1996},  possibly related to the strong clustering of voids 
and their primordial precursors \citep{Abbas2007}.

We have undertaken a new multi-wavelength Void Galaxy Survey (VGS) of $\sim$60 geometrically selected void galaxies.  In this project we intend to study in detail the gas content, star formation history and stellar content, as well as the kinematics and dynamics of void galaxies and their companions in a broad sample of void environments.  
Each of the galaxies has been selected from the deepest interior regions of identified voids in the SDSS redshift survey on the basis of a unique  geometric technique, described in Section~\ref{sec:select}, with no a priori selection on intrinsic properties of the void
  galaxies (ie. luminosity or color).

We present here the results of a pilot study,  observing 15 void galaxies  in H \textsc{i} at the Westerbork Synthesis Radio Telescope (WSRT). 
Our selection procedure for this pilot sample and for the full VGS are explained in Section~\ref{sec:select}.
Our observations, detailed in Section~\ref{sec:obs}, have sufficient sensitivity and resolution to allow us to map the gas distribution and kinematics for each of our target galaxies.  
In addition, the large field of view and wide redshift range allows a simultaneous detection of faint nearby companions, as well as background galaxies residing in higher density regions. 
This allows a very accurate determination of the local environment and direct evidence of gas interactions.
A detailed presentation of the results is given in Section~\ref{sec:results}.  An extensive analysis of the void galaxy properties is the subject of Section~\ref{sec:analysis}.
 A discussion of our preliminary findings and speculations as to their implications are to be found in Section~\ref{sec:speculation}, with conclusions presented in Section~\ref{sec:conclusion}. Throughout the paper we have assumed H$_\textrm{o}=70\,{\rm km}\,{\rm s}^{-1}\,{\rm Mpc}^{-1}$, except where noted.

\section{Sample Selection} 
\label{sec:select}
Voids are the (mostly) empty regions between the filaments, walls and clusters in the 
Cosmic Web. While these large underdense regions represent one of the most prominent aspects of 
the cosmic matter distribution, there is no unanimity with respect to their definition. Unlike e.g. clusters 
of galaxies, voids are not well-defined physical objects. Instead, they are the regions surrounding the minima 
in the cosmic matter and galaxy distribution. Different opinions exist on their extent and boundary, 
the medium with respect to which they should be identified, and the practical implementation 
of such definitions to the galaxy distribution. As a result, there is a large variety of void finding formalisms. Some refer to the galaxy distribution, while others use
the dark matter density field. Some identify isolated spherical voids, while others attempt to reconstruct the nontrivial 
void shapes and geometries by means of overlapping spheres or cubes. While all methods can generally identify an underdense volume, they differ significantly in identifying the location of the edge of 
the void. A telling illustration of this can be found in the overview and quantitative comparison of various 
void finding algorithms by \cite{Colberg2008}. 

The challenge for studies of voids is therefore to identify them in an unbiased and cleanly defined manner. For our 
purpose it is of key importance that we make no a priori assumptions about the scale and shape of voids, and that we 
find voids entirely independent of the intrinsic properties of galaxies. To this end, we invoke a unique geometric void finding algorithm. 
It is purely based on the local spatial structure of the galaxy distribution and guarantees the definition of an 
unbiased sample of void galaxies.

\subsection{Geometric Void Identification}
\label{sec:voidfind}

Our galaxy sample is defined on the basis of the optical SDSS redshift survey, which covers over 10,000 square degrees and catalogs redshifts for over 900,000 galaxies  brighter than 17.77 Petrosian magnitudes in the $r$-band and more than 55$^{\prime\prime}$ away from any other cataloged galaxy.
The pilot program void galaxies were selected 
from the SDSS Data Release 3 (DR3), while the full Void Galaxy Survey uses the complete DR7.

The first step in detecting void galaxies is outlining the voids in the survey volume. It is important for our purpose that 
we follow a strictly geometric procedure. This involves the reconstruction of the density field from the spatial galaxy distribution, 
followed by the identification of void regions within the spatial density field. Finally, we search for the SDSS galaxies 
which lie within the interior of the identified voids. 

\subsubsection{The DTFE density field}
\label{sec:dtfe}
For the density field reconstruction we use the DTFE procedure, the Delaunay Tessellation Field Estimator 
\citep{Schaap2000,Schaap2007,weyschaap2009}. This technique translates the spatial distribution of the galaxies in 
the SDSS, in a volume from $z=0.003$ to $z=0.03$, into a continuous density field. In addition to the computational 
efficiency of the procedure, the density maps produced by DTFE have the virtue of retaining the anisotropic and 
hierarchical structures which are so characteristic of the Cosmic Web \citep{Schaap2007,weyschaap2009}. The recent 
in-depth analysis by \cite{Platen2010} has shown that for very large point samples, the DTFE even outperforms more 
elaborate high-order methods with respect to quantitative and statistical evaluations of the density field. As a 
result, the DTFE density field is highly suited for objectively tracing structural features such as walls, filaments 
and voids. 

The DTFE procedure involves four steps. The first step is the definition of the galaxy sample and 
corresponding survey volume. The subsequent step consists of the computation of the Delaunay tessellation 
defined by the spatial galaxy distribution. DTFE exploits the adaptivity of this tessellation to the local density and 
geometry of the generating spatial point process. The third step involves the local density estimate at each of the 
sample galaxies' positions. This is based on the volume of the contiguous Voronoi cell, the volume defined by all Delaunay 
tetrahedra to which a given sample galaxy belongs. In the final step, these local density values form the basis for a 
piecewise linear interpolation within the volume of each Delaunay tetrahedron. In this, the Delaunay Tessellation Field 
Estimator is the multidimensional equivalent of simple piecewise one-dimensional linear interpolation from an irregularly 
distributed set of points which uses the Delaunay triangulation as an adaptive and irregular interpolation grid.

The resulting product of the DTFE procedure is a volume-covering continuous density field. For a an extensive 
description of the full DTFE procedure we refer to \cite{weyschaap2009}. Below we shortly describe the 
key steps of the DTFE machinery. 

\bigskip
\noindent {\it SDSS DR3 sample}\\
The SDSS DR3 galaxy sample is a {\it magnitude} limited sample, consisting of 
galaxies brighter than $m_r=17.77$. For our density field determination we take all 
sampled information, which makes it necessary to correct for the inhomogeneous selection 
process. By default, we assume that all galaxies - independent of their luminosity - are a 
fair tracer of the underlying galaxy density field. 

Following this assumption, we correct for the dilution as a function of survey depth by weighing 
each sample galaxy by the reciprocal $w(z)$ of the radial selection function $\psi(z)$ at the 
distance of the galaxy.  For the SDSS, the selection function $\psi(z)$ as a function of redshift $z$ 
is well fitted by the expression forwarded by \cite{Efstathiou2001}
\begin{equation}
\psi(z)\,=\,\exp\left\{-\left(\frac{z}{z_r}\right)^{\beta}\right\}\,,
\label{eq:psiz}
\end{equation}
where $z_r$ is the characteristic redshift of the distribution and $\beta$ specifies 
the steepness of the curve. 

\bigskip
\noindent {\it Delaunay Tessellation}\\
The {\bf Delaunay Tessellation} of the galaxy distribution divides up the sample volume in a unique 
volume-covering tiling of tetrahedra. Each of these Delaunay tetrahedra is defined by a set of four galaxies 
in such a way that their circumscribing spheres does not contain any of the other galaxies of the generating 
galaxy sample \citep[][]{Delaunay1934}. For practical purposes, we assume {\it vacuum boundary conditions}: outside 
the SDSS galaxy sample volume we take the minimal assumption of having no galaxies. For its efficient 
computation we use the CGAL library \footnote{CGAL is a \texttt{C++} library of algorithms and data 
structures for Computational Geometry, see \url{www.cgal.org}}. 

\bigskip
\noindent {\it DTFE Density Estimate}\\
The DTFE density value estimate at the location of each galaxy is taken to be inversely proportional 
to the volume of the {\it contiguous Voronoi cell}, i.e. the region defined by all Delaunay tetrahedra 
of which a given galaxy is a vertex. For a sample galaxy $i$, we identify all $N_i$ neighbouring 
Delaunay tetrahedra $\mathcal{T}_j$,  which together constitute the contiguous Voronoi cell ${\cal W}_i \cup_j \mathcal{T}_j$. 
Summation of the individual tetrahedral volumes $V({\cal T}_j)$ yields the volume of the contiguous Voronoi cell, 
\begin{equation}
V({\cal W}_i)\,=\,\sum_{j=1}^{N_i}\,V({\cal T}_j)\,.
\end{equation}
The resulting DTFE estimate $\widehat{f_i}$ of the density field at sample point $i$ is 
\begin{equation}
  \widehat{f_i}\,=\,\frac{4\,w(z_i)}{V({\cal W}_i)}\,.
\label{eq:densdtfe}
\end{equation}
where the weight $w(z_i)$ is the sample selection weight at the galaxies' redshift $z_i$. Note that the 
factor of four takes account of the fact that in three dimensions each sample point belongs to four tetrahedra. 
In practice, the density of all galaxies is calculated by looping in sequence over all Delaunay tetrahedra. 

\bigskip
\noindent {\it DTFE Density Field Interpolation}\\
DTFE uses the adaptive and minimum triangulation properties of Delaunay tessellations to use the tessellations as adaptive spatial 
interpolation intervals for irregular point distributions \citep{Bernardeau1996}. By doing so, the DTFE generalizes the concept 
of a natural interpolation interval to any dimension $D$. 

Once the Delaunay tessellation has been constructed, and the densities at each sample point determined, the DTFE determines 
the density gradient within each Delaunay tetrahedron $\mathcal{T}_j$. The gradient can be directly inferred 
from the density values at its four vertices, i.e. at the location of the four sample galaxies defining the 
tetrahedron. 

Following the determination of the density gradients in all Delaunay tetrahedra, the DTFE density value at any point 
${\bf \widehat{r}}$ within the sample volume can be calculated by determining in which tetrahedron it is located and 
subsequently computing its density estimate from the simple linear interpolation equation. To obtain an image of the 
density field, the density is calculated at each of the voxel locations of the image grid.  

\subsubsection{The Watershed Void Finder}
\label{sec:wvf}
The Watershed Void Finder (WVF, \citealt{Platen2007}) is applied to the DTFE density field for identifying its underdense void basins. 
The WVF is an implementation of the {\it Watershed Transform} for segmentation of images of the galaxy and matter distribution 
into distinct regions and objects and the subsequent identification of voids. 

The basic idea behind the watershed transform finds its origin in geophysics. It delineates the boundaries of the separate domains, 
the {\it basins}, into which yields of e.g. rainfall will collect. The analogy with the cosmological context is straightforward. 
By equating the cosmic density field to a landscape, the WVF identifies the valleys with {\it cosmic voids}. They are separated 
from each other by boundaries consisting of the {\it walls} and {\it filaments} in the Cosmic Web. These are identified with the 
related Spine formalism \citep{Aragon2008}.

The WVF consists of eight steps, which are extensively outlined in \cite{Platen2007}. For the success of the WVF it is of the utmost 
importance that the density field retains its topology. To this end, the two essential first steps relate 
directly to DTFE, which guarantees the correct representation of the hierarchical nature, the weblike morphology dominated by 
filaments and walls, and the presence of voids \citep{weyschaap2009}. 
 Because in and around low-density void regions the raw density field
 is characterized by a relatively high level of noise, a second essential
 step suppresses the noise using a R$_f$= 1 $h^{-1}$ Mpc Gaussian filter. The
 subsequent central step of the WVF formalism
 consists of the application of the discrete watershed transform on this
 filtered density field. Thus providing us with a set of voids. Analysis
 of mock catalogs show that a choice of R$_f$ = 1 $h^{-1}$ Mpc recovers the
majority
 of small scale voids, and within
 distances of 100 $h^{-1}$ Mpc errors are dominated by the inherent
limitations
 of the observational redshift survey \citep{Platen2009}.

Unlike most void finders, the WVF has the advantage of having no a priori assumptions about scale and shape of 
voids \citep[see][ for a comparison of different algorithms]{Colberg2008}. Instead, these are completely determined by the topological 
structure of the cosmic density field. Because it identifies a void segment on the basis of the crests in a density field surrounding a 
density minimum, it is able to trace the void boundary even though it has a distorted and twisted shape. Also, because the contours around 
well chosen minima are by definition closed, the transform is not sensitive to local protrusions between two adjacent voids.

\begin{figure}
\centering
\includegraphics[width=0.99\textwidth]{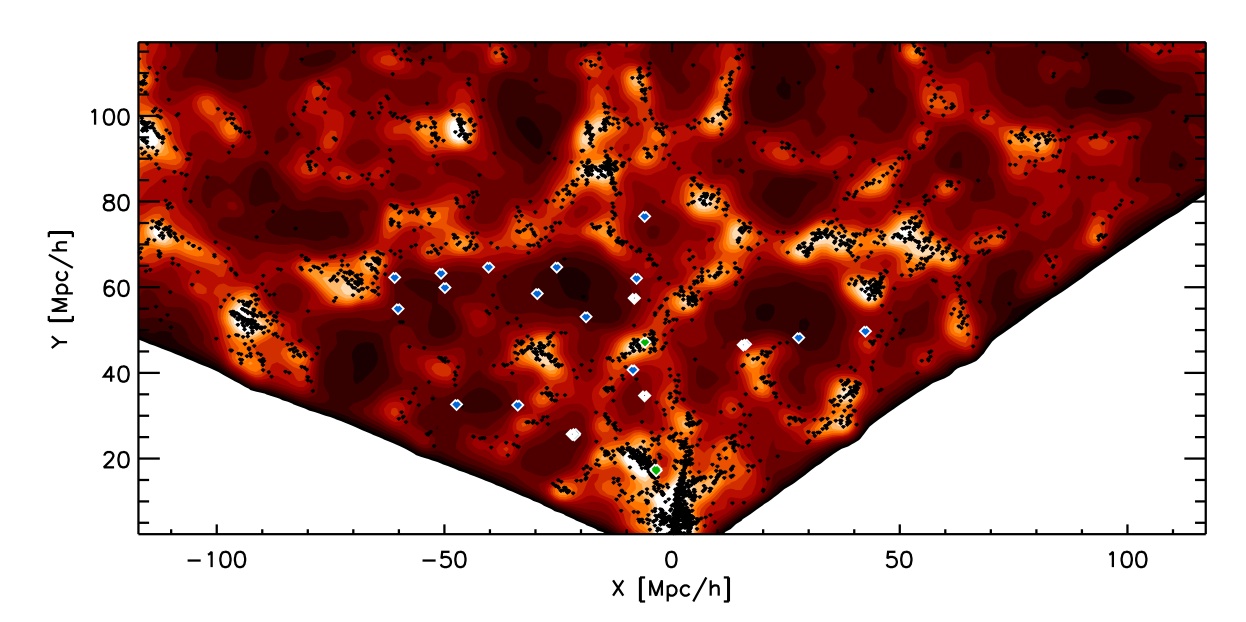}
\caption{SDSS density map and galaxies in the SDSS galaxy
  redshift survey region from which we selected the galaxies in the
  Void Galaxy Survey, in a slice of thickness 4$h^{-1}$Mpc. The DTFE computed galaxy density
  map, Gaussian smoothed on a scale of $R_f=1h^{-1}$Mpc, is represented by
  the colorscale map. The map runs from dark red, at lowest void densities, to beige, the average cosmic galaxy density.
  The SDSS galaxies are superimposed as dark dots. The pilot sample
  void galaxies are represented by white diamonds, while the blue
  diamonds indicate the position of void galaxies from the full Void Galaxy Survey.
  The green diamonds are control sample galaxies (see Section~\ref{sec:control}).
\label{fig:sdssmapvoid}}
\end{figure}

A representative DTFE density map in a slice through the SDSS galaxy distribution is shown in Figure~\ref{fig:sdssmapvoid}, 
with superimposed symbols indicating the original spatial distribution of galaxies and subsequently determined void galaxies. 

\subsection{Void Galaxy Selection}
\label{sec:galselect}

Upon having obtained the complete list of voids in the SDSS survey volume, and the spectroscopically targeted void galaxies within their realm, we 
evaluate for each of the galaxies whether it conforms to a set of additional criteria. The galaxy should be 
\begin{itemize}
\item[$\bullet$] located in the interior central region of clearly defined voids, 
if possible near the center, and be as far removed from the boundary of the voids as possible.  
\item[$\bullet$] removed from the edge of the SDSS survey volume, as we do not wish to have galaxies 
in voids which extend past the edge of the SDSS coverage. 
\item[$\bullet$] not within $\approx 750$km s$^{-1}$ from a foreground or background cluster. This 
assures that the presence in a void of a galaxy can not be attributed to the Finger of God effect. 
\item[$\bullet$] preferably within a redshift $0.01<z<0.02$, allowing sufficient sensitivity and 
resolution in our observations of the gas structure and kinematics in the galaxies.
\end{itemize}
Other than the SDSS redshift survey limit of 
\begin{equation}
M_r = 17.77 - 5 \log_{10} \frac{c z}{H_o~ 10pc} 
\end{equation}
there is no selection on luminosity or color of the void galaxies in our sample, 
the selection is purely dictated by the local geometric structure of the
  Cosmic Web. 
  
The spectroscopic selection criteria of the SDSS do exclude bright, nearby galaxies and omit some galaxies in particularly crowded fields, however these effects are not relevant to our sample.  There is also a poorly constrained surface brightness selection in the SDSS main galaxy sample.

Apart from avoiding the Finger of God effect, 
note that redshift distortions due to cosmic large scale flows, and
as a result of cluster infall, are not affecting our selection. These
tend to amplify the density and contrast of filaments, sheets and
cluster outskirts in redshift space, while leading to emptier voids:
void galaxies detected in redshift space are therefore highly likely
to also live in physical voids \citep[cf.][]{kaiser1987,praton1997,shandarin2009}.

\begin{figure}
\centering
\setlength\fboxsep{1pt}
\setlength\fboxrule{0.5pt}
\fbox{\includegraphics[width=0.95\textwidth]{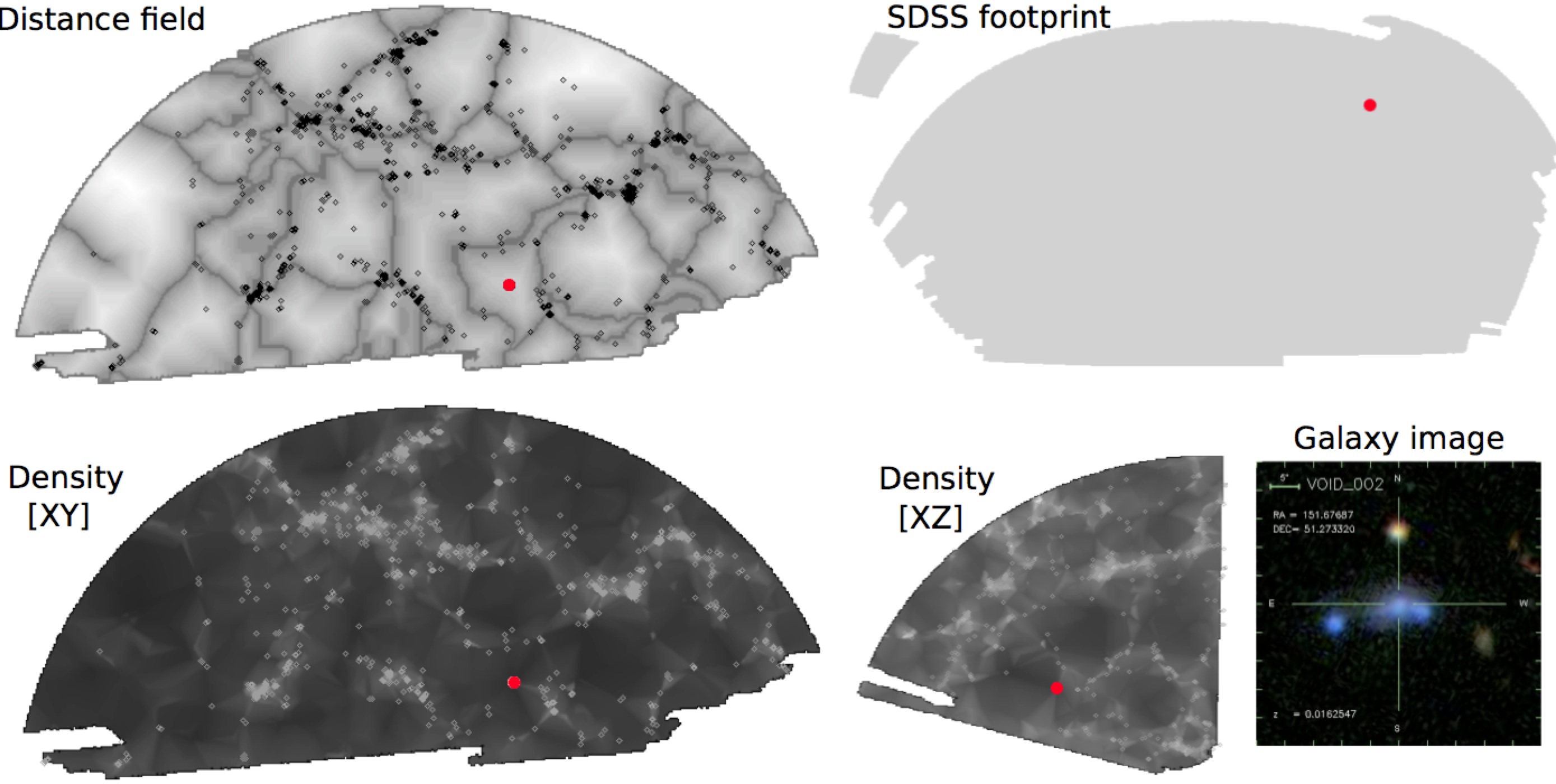}}
\fbox{\includegraphics[width=0.95\textwidth]{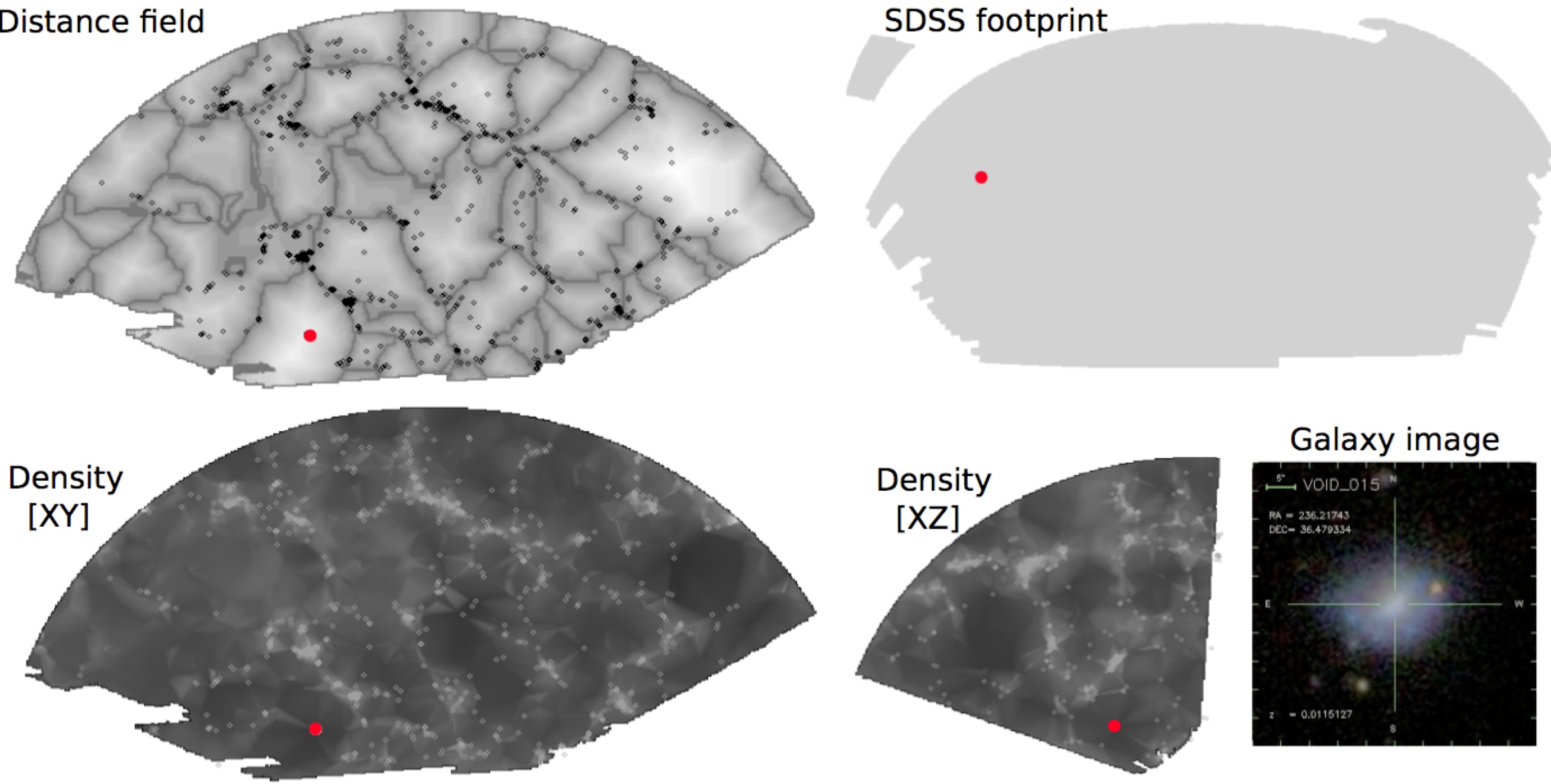}}
\caption{Two examples of selecting our galaxies by geometry from the SDSS using different visualizations of the density field. Top right 
panel: SDSS DR7 sky map footprint, with the void galaxy indicated by a heavy red dot, to show it is positioned away from the survey edge. Bottom right: galaxy image from SDSS 
database. Bottom left and center: DTFE density greyscale maps in two mutually perpendicular slices intersecting at the 
galaxy location. Top left: The SpineWeb/watershed contours of the density field are shown in dark gray on top of the distance field (described in Section~\ref{sec:galselect}) in gray scale. 
The locations of the SDSS DR3 galaxies within the thin density field and distance field slices are indicated by the diamond shaped points, 
while the heavy red dot represents the target galaxy located away from any walls or filaments and deep within the underdensity. }
\label{fig:select}
\end{figure}

Figure~\ref{fig:select} illustrates our selection procedure for two cases. The locations of the void galaxy in the SDSS footprint (heavy dot, top right) are removed from the survey edge and each other.
The DTFE density map in two mutually perpendicular slices intersecting at the galaxy location, along the line pointing radially towards the 
observer (bottom left and center), show  their position centrally located in the deepest underdensities. To accentuate this, the images also contain a map of the spinal/watershed contours of the density field overlaid on top of 
the distance field in order to give an idea of the spatial structure of the surrounding large scale structure (top left). The distance field shows the minimum distance from the void edge, and is maximized at the void center. The 
locations of the SDSS DR3 galaxies within these thin slices are indicated by the diamond shaped points, while the 
heavy dot represents the target galaxy. A small galaxy image from the SDSS database (bottom right) shows the optical appearance of our geometrically selected void galaxy.

\subsection{Resulting Sample}
\label{sec:sample}

\begin{figure}
\centering
\includegraphics[width=0.99\textwidth]{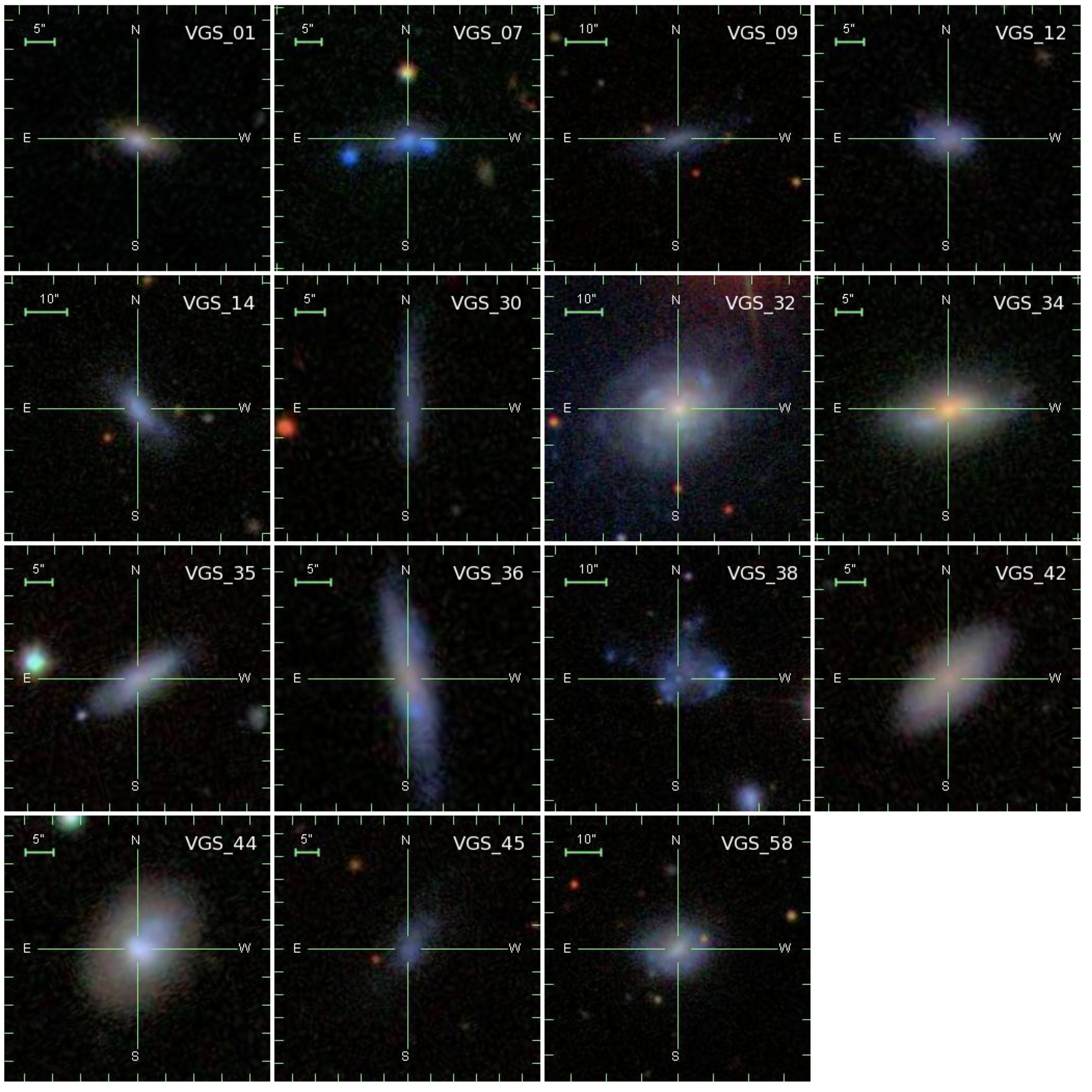}
\caption{Our sample of void galaxies, scaled to the same physical size. Composite color images are taken from the online 
SDSS Finding Chart tool. } 
\label{fig:poststamps}
\end{figure}

In Figure~\ref{fig:poststamps} we present the SDSS color images of our 15 selected void galaxies, taken from the SDSS online 
Finding Chart tool (DR7, \citealt{Abazajian2009}). The images have been scaled to the same physical scale. The parameters for the target galaxies, extracted from the SDSS catalog, are given 
in Table~\ref{tab:params} along with the program and SDSS names. In the remainder, we 
will mainly use the program name ``VGS\_$\alpha$'', unless otherwise stated. Likewise, the companion galaxies found in 
our program will receive a program name ``VGS\_$\alpha$a'', ``VGS\_$\alpha$b'', etc., where $\alpha$ is the 
number for the parent void galaxy. 

\begin{figure}
\centering
\includegraphics[width=0.95\textwidth]{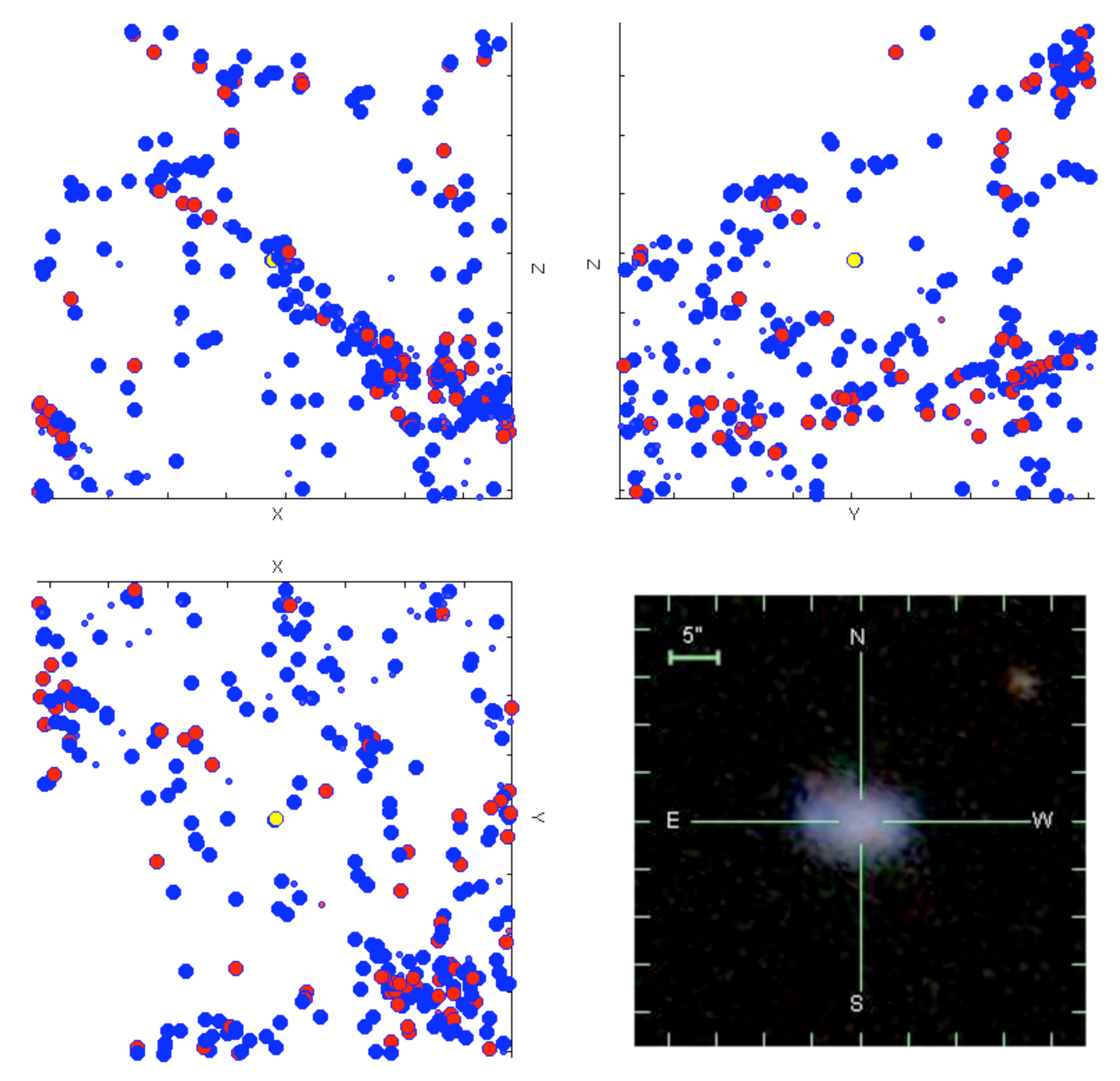}
\caption{The SDSS galaxy distribution in and around the void in which galaxy VGS\_12 of our 
void galaxy sample is located, seen from three different perspectives. The galaxies are located 
within a box of size 24 $h^{-1}$ Mpc centered on VGS\_12 (in yellow). Bright galaxies have large 
dots (B~$<-16$), fainter ones are depicted by smaller dots. Redder galaxies, with $g-r>0.6$ are 
indicated by red dots. Bluer galaxies, with $g-r\le0.6$, are indicated by blue dots. The SDSS image of the galaxy is shown in the bottom right panel.}
\label{fig:voidgalspatial3}
\end{figure}

\begin{figure}
\centering
\includegraphics[width=0.95\textwidth]{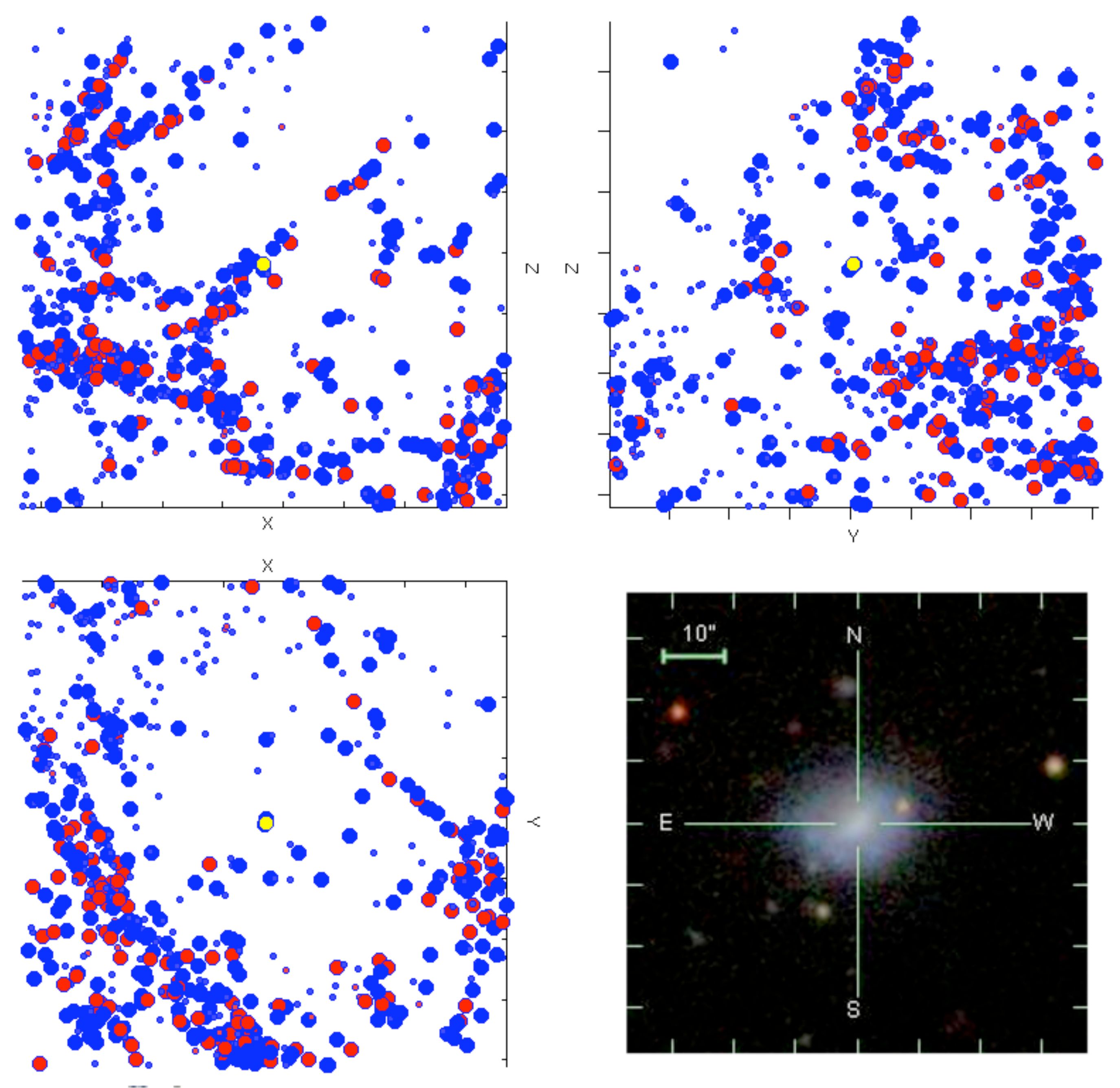}
\caption{The SDSS galaxy distribution in and around the void in which galaxy VGS\_58 of our 
void galaxy sample is located, seen from three different perspectives. The galaxies are located 
within a box of size 24 $h^{-1}$ Mpc centered on VGS\_58 (in yellow). Bright galaxies have large dots (B~$<-16$), fainter 
ones are depicted by smaller dots. Redder galaxies, with $g-r>0.6$ are indicated by red dots. Bluer galaxies, with $g-r\le0.6$, are indicated by blue dots. The SDSS image 
of the galaxy is shown in the bottom right panel.}
\label{fig:voidgalspatial14}
\end{figure}

To obtain an impression of the true thee-dimensional environment of the selected void galaxies, we have 
included a set of 3-D galaxy distribution visualizations around void galaxies VGS\_12 (Figure~\ref{fig:voidgalspatial3}) 
and VGS\_58 (Figure~\ref{fig:voidgalspatial14}). In a box of size 24 $h^{-1}$ Mpc centered around 
each of these void galaxies, the structure of the surrounding galaxy distribution is depicted from the three 
different perspectives. In these panels, bright galaxies have large dots (B~$<-16$), while fainter ones are depicted by 
smaller dots. Redder galaxies, with $g-r>0.6$ are indicated by red dots, and bluer galaxies, with $g-r\le0.6$, are indicated by blue dots. 

Void 14 lies well within the interior of a huge void, surrounded by filamentary structures. On one side these are embedded 
within a massive wall-like complex. The 
geometry of the spatial galaxy distribution around the polar ring galaxy VGS\_12 has been discussed at length in 
\cite{Stanonik2009}: the galaxy lies within a tenuous wall in between two voids. This is perhaps most evident 
in the top left panel, where we see a planar concentration forming the division between a large void (to the 
top of the box) and a smaller void (to the bottom of the box). A comparison between the spatial 
distribution of red and blue dots reveals a strong segregation in the low-density regions. In particular the 
VGS\_58 maps illustrate the avoidance by red galaxies of underdense areas. A similar distinction in spatial 
distribution between faint and bright galaxies is far harder to notice, if at all.

\begin{figure}
\centering
\includegraphics[width=4in,angle=90]{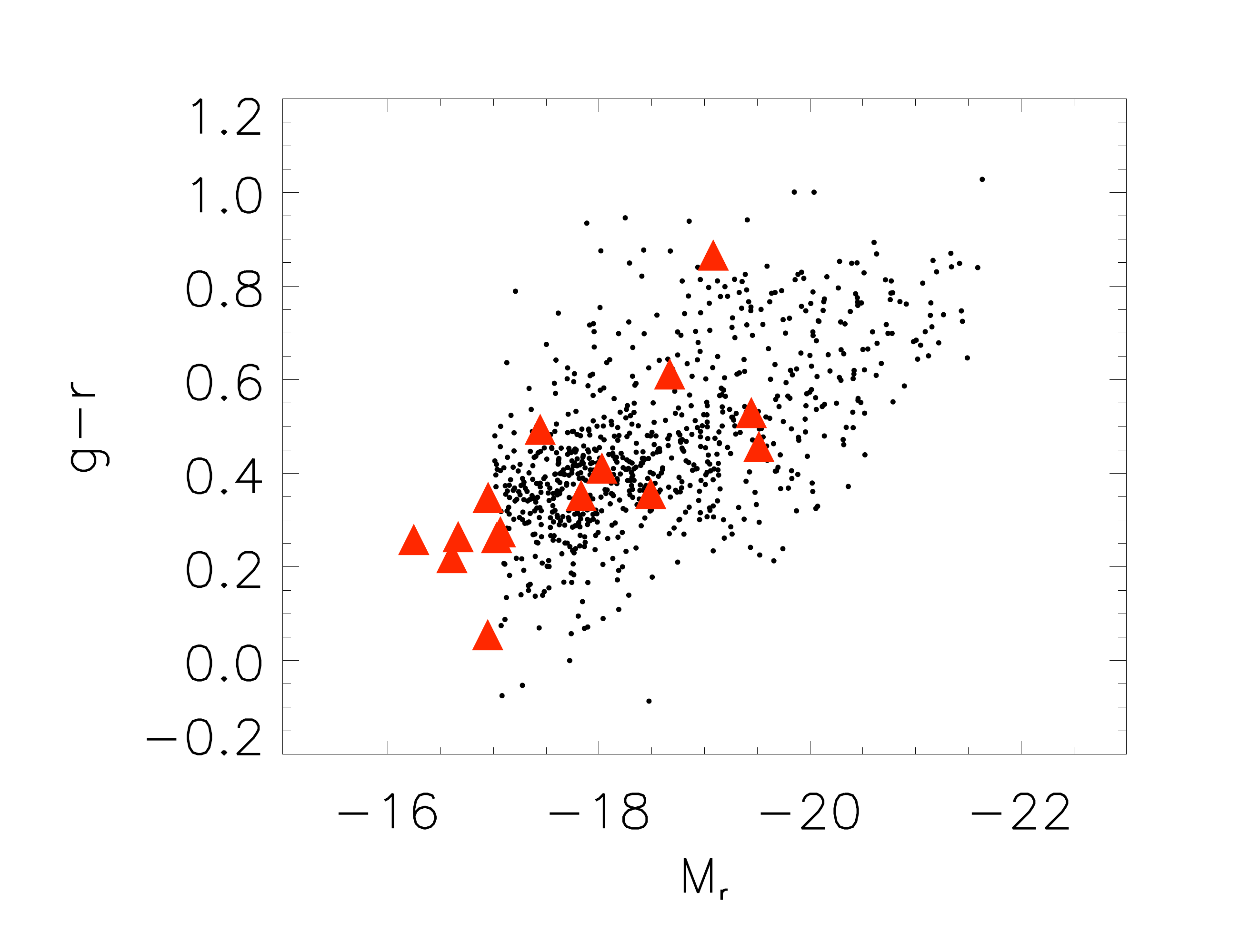}
\caption{ Color-magnitude diagram for our void galaxy sample (triangles), 
compared with a volume limited ($z<0.02$, M$_r<-16.9$) sample of  void galaxies from the SDSS. 
\label{fig:colmag}  }
\end{figure}

The nature of our galaxies can be immediately appreciated from the color-magnitude diagram in Figure~\ref{fig:colmag}. For 
comparison we include a volume-limited sample of galaxies in underdense regions defined by our void-finding algorithm from the SDSS DR7, including all galaxies brighter than 
$M_r=-16.9$ within a redshift range $z<0.02$. Most of our galaxies are in the blue 
cloud and at the faint end of the galaxy luminosity function, where the bulk of SDSS void galaxies reside. See Section \ref{sec:analysis} for a detailed discussion of how our void galaxy sample compares with other void galaxies samples.

Table~\ref{tab:voidparams} contains parameters for the voids surrounding each target galaxy, taken from our void finding algorithm applied to the SDSS reconstructed density field.  For the galaxies that seem to reside in between two voids we have quoted two values, for the two nearest voids.\\
\ \\
\textit{Column} (1) lists the program name.\\
\textit{Column} (2) lists the (equivalent) void radius $R_{void}$,\\
\begin{equation}
R_{void}\,\equiv\,\left({\displaystyle 3\, V_{void} \over \displaystyle 4 \pi}\right)^{1/3}\,,
\end{equation}
where $V_{void}$ is the volume of the void. It corresponds 
to the radius the void would have if it were spherical. \\
\textit{Column} (3) gives $D_{gal,void}$, the distance of the galaxy from the void center.\\
\hskip 0.5truecm \textit{Column} (4) gives the the ratio of void center distance to void radius, \\
\begin{equation}
DR\,\equiv\,{\displaystyle D_{gal,void} \over \displaystyle R_{void}}.
\end{equation}
Note that nearly all detected voids are elongated, so that galaxies located within the voids' 
interior may still have $DR > 1$.  \\
\textit{Column} (5) provides  the filtered density contrast of the void, \\
\begin{equation}
\delta\,\equiv\,{\displaystyle \rho_{void} \over \displaystyle \rho_u}\,-1\,\,, 
\end{equation}
at a filtering scale R$_f=1$h$^{-1}$Mpc. Here, $\rho_u$ is the mean density, estimated by the
average point intensity from the SDSS volume (see \citealt{Platen2009}). \\
\textit{Column} (6) lists the distance to the nearest neighbor.\\
\textit{Column} (7) lists the average distance of the six nearest neighbors.\\
\textit{Column} (8) gives  the void name, taken from \cite{Fairall1998} and \cite{Hoyle2002a}.\\

While all targets were ideally chosen for their isolation and central location within the void, our pilot sample selection was limited to the galaxy 
redshift survey data available with the incomplete DR3.  As the catalog has been completed since then some of these targets have proven to be in 
slightly less centrally located, though still underdense, environments. This, and the fact that most voids have rather elongated shapes, 
explains why several of the sample galaxies have a distance $R_{void}$ to the void center which is in the order of the (equivalent) void 
scale $R_{void}$. Galaxies VGS\_07, VGS\_34, VGS\_44, and VGS\_58 are the best in terms of location near the center of the void and not near a wall-like 
configuration.

\section{Observations} 
\label{sec:obs}
Our targets were observed 2006-2007 with the Westerbork Synthesis Radio Telescope (WSRT) in Maxi-short configuration.  
The shortest baselines of 36, 54, 72 and 90 meters optimize surface brightness sensitivity in a single 12 hour observation. Its longest spacing of 2754 meters results in an angular resolution of approximately 20$^{\prime\prime}$.
  Observing parameters are detailed in Table~\ref{tab:obsparams}.  We observed using 512 channels per 10 MHz bandwidth, at four simultaneously observed frequencies, each with two polarizations.  Along with the observation centered at the target galaxy, three other simultaneous observations were made centered at frequencies 8.3 MHz, -8.3 MHz and -16.6 MHz removed from the target galaxy, resulting in a velocity coverage of nearly 10,000 km s$^{-1}$ with each telescope pointing.  Target redshifts ranged from 0.012 to 0.022, so channel increments of 19.5312 kHz give a typical velocity resolution of 4.25 km s$^{-1}$ using the heliocentric optical velocity definition.  After applying Hanning smoothing this is reduced to 8.6 km s$^{-1}$. The full width at half maximum of the primary beam is about 36$^{\prime}$, which at a typical redshift of 0.015 is a 700 kpc $\times$ 700 kpc area on the sky.

Each observation consisted of a 12 hour target exposure, calibrated by two 15 min snapshots, one observed before and one observed after, using two of 3C48, 3C286, 3C147 and CTD93.  Due to the location's impressive phase stability, our calibration observations once every 12 hours are sufficient for phase calibration.  The absolute flux scale is determined using the flux scale of \cite{Baars1977} in the AIPS task SETJY.  At a first channel frequency of 1406.9212 MHz, a corresponding flux of 14.79 Jy, 16.00 Jy, and 22.10 Jy are used for 3C286, 3C48, and 3C147, respectively. For the non-standard calibrator CTD93 we use the 20 cm flux of 4.83 Jy, as listed in the VLA calibrator manual.  All calibration is done using standard AIPS procedures. 

Continuum emission is subtracted from the UV data by interpolation from the line free channels.  Image cubes are created with a CLEAN box around the H \textsc{i} emission that cleans down to 0.5 mJy beam$^{-1}$ (1 $\sigma$). Images are created with a robust parameter of 1 \citep{Briggs1995} and corrected for the primary beam.   Data cubes are then Hanning smoothed to achieve a sensitivity of 0.4 mJy beam$^{-1}$.  Zeroth and first moment maps are made over a narrow velocity range encompassing the detection, with Hanning and Gaussian smoothing over 25 km s$^{-1}$ and 30$^{\prime\prime}$ and 1.5$\sigma$ clipping.  For galaxies removed from the beam center the cutoff value is scaled to correct for the primary beam. 

The continuum baseline fit from the line free channels, which is used to determine the continuum subtraction in the UV data, is also used to create a 1.4 GHz continuum image.  Images are created with a robust parameter of 1 and CLEANed down to 0.2 mJy beam$^{-1}$ (1.5 $\sigma$).  

The separation between a void galaxy and the edge of the neighboring velocity range is about 5 MHz once edge channels are excluded, which at the average distance of 70 Mpc corresponds to a cosmological distance of $\sim$15 Mpc.  This is large enough to distinguish these neighboring velocity ranges as removed from the target void, which maximally extend to about 15 Mpc in radius (Table \ref{tab:voidparams}).
For blind H \textsc{i} detections throughout the neighboring velocity range, excluding the target galaxy, data cubes are smoothed to 20 km s$^{-1}$ with a 6 kilowavelength taper, producing a 40$^{\prime\prime}$ beam and 1$\sigma$ sensitivity of 0.25 mJy beam$^{-1}$. Zeroth moment maps are made with Hanning and Gaussian smoothing and 5$\sigma$ clipping before correcting for the primary beam to provide a simple blind detection algorithm.  We then selected by eye the regions with HI emission that are extended
both spatially and in velocity and call this our control sample.

\section{Results} 
\label{sec:results}
In our sample of 15 void galaxies (Table~\ref{tab:params} and Figure~\ref{fig:vgs}) we have one non-detection, and discovered one previously known and five previously unknown companions (Table~\ref{tab:company} and Figure~\ref{fig:comps}).  Of the five void galaxies with companions, two are interacting in H \textsc{i}.  All H \textsc{i}-detected companions have optical counterparts within the SDSS.  
Of the nine isolated void galaxies, two exhibit clear irregularities in the kinematics of their gas disks.  Target galaxies have a range of H \textsc{i} masses from $3.5 \times 10^8 M_{\odot}$ to $3.8 \times 10^9 M_{\odot}$  and one non-detection with a 3$\sigma$ upper limit of $2.1 \times 10^8 M_\odot$, assuming a 150 km s$^{-1}$ velocity width.  Companion galaxies have masses ranging from $5.0 \times 10^7 M_{\odot}$ to $4.5 \times 10^8 M_{\odot}$.  H \textsc{i} properties of the target and companion galaxies are given in Table~\ref{tab:vgs}.  

\subsection{HI Data}
\label{sec:hidata}
H \textsc{i} mass is calculated using the standard formula, $M_{\textrm{H \textsc{I}}}=2.356 \times 10^5 \: $d$^2 \int $S$ \: dv \; M_\odot$, where d is the distance in Mpc, S is in Jy and \textit{dv} is in km s$^{-1}$. The integrated flux is taken from the total intensity maps described in Section~\ref{sec:obs}.  A rough global profile is constructed by summing the flux in a tight box surrounding the target.  The 50\% H \textsc{i} line width ($W_{50}$)  and  20\% H \textsc{i} line width ($W_{20}$) are corrected for instrumental broadening following \cite{Verheijen2001}, and broadening due to random motions is neglected. Errors of 15 km s$^{-1}$ reflect uncertainties of one channel on either side. Systemic velocities are estimated from tilted ring fits of the velocity field, with typical errors of 5 km s$^{-1}$.

\begin{figure}
\centering
\includegraphics[width=6in]{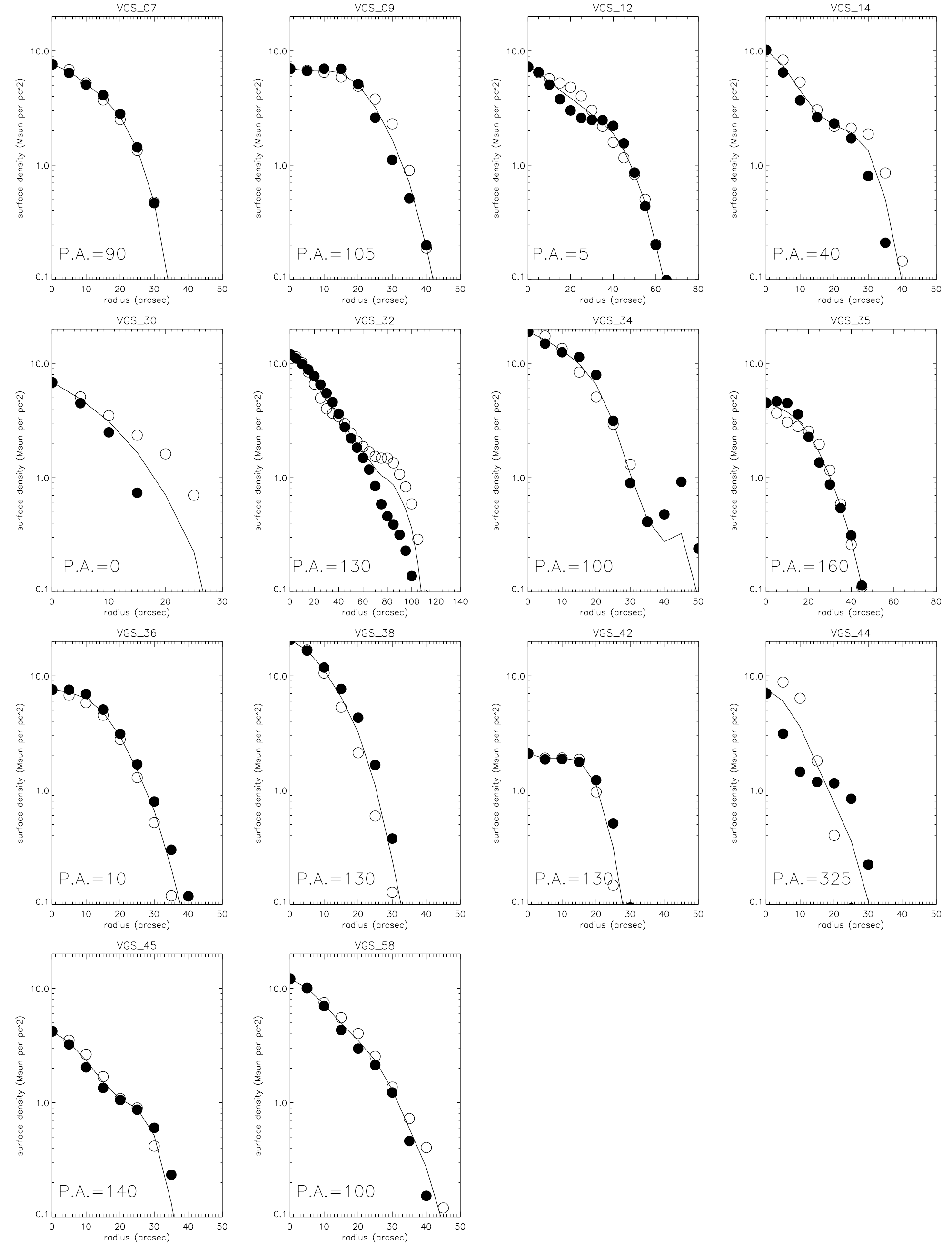}
\caption{Radial surface density profiles for the east (open circles) and west (filled circles) sides of each disk, with the average overdrawn.  \label{fig:surfdens}}
\end{figure}

Because the majority of the void galaxies are not well enough resolved to reliably fit tilted rings to the kinematics of the gas disk, we measure the radial surface density profile following the iterative deconvolution method described by \cite{Lucy1974} as implemented by \cite{Warmels1988}.  This technique projects the two dimensional surface brightness map along the disk minor axis to create a one dimensional strip distribution which shows projected surface brightness as a function of position along the disk major axis.  Assuming an azimuthally symmetric and circular disk, this strip distribution is then iteratively modeled to reconstruct the radial surface density profile (Figure~\ref{fig:surfdens}).  
As discussed by \cite{Swaters2002}, this iterative deconvolution method is known to restore too much flux to the disk center, so while the iterative fit was stopped at 10 iterations to prevent this, we may systematically be slightly underestimating our radii and significantly overestimating our central surface densities. 
Position angles are determined by considering both the $g$-band optical as well as the H \textsc{i} kinematic major axes.  The H \textsc{i} radius, R$_{\textrm{H \textsc{I}}}$, is taken as that point where the surface density drops below 1 $M_\odot$ pc$^{-2}$.  Upper limits are listed for those galaxies with an H \textsc{i} extent smaller than the synthesized beam size. 
Typical uncertainty in the radius is 5$^{\prime\prime}$, about 1.5 - 2 kpc for our void sample.  This is much larger than the typical 0.5$^{\prime\prime}$ (150 - 200 pc) errors in the optical r$_{90}$ radius, containing 90\% of the Petrosian flux.

As all detected target galaxies exhibit disklike rotation, the inclination corrected circular velocity is used to estimate a dynamical mass, $M_{dyn}$, for the volume interior to R$_{\textrm{H \textsc{I}}}$ as 
\begin{equation}
M_{dyn}(r < R) = \frac{W_{50}^2 R_{\textrm{H \textsc{I}}}}{\sin^2 i ~G} \; M_{\odot} .
\end{equation}
Inclinations, $i$ are calculated from SDSS $r$-band isophotal major and minor axes. Following the assumption that disks be intrinsically oblate and axisymmetric, with a three dimensional axis ratio of q$_o$ = 0.19 \citep{Geha2006}, we use
\begin{equation}
\sin i = \sqrt{\frac{1-(b/a)^2}{1-q^2_o}} .
\end{equation}
A choice of q$_o$ = 0.3 changes the resulting inclinations by less than 10\%.  For targets which have no cataloged $r$-band major and minor axes, we use $u$-band values. The $r$-band and $u$-band inclinations are generally in agreement to within 10\%. Due to these uncertainties in the inclination and in the H \textsc{i} radius we expect our dynamical masses to be accurate within roughly a factor of two.

\subsection{Radio Continuum Data}
\label{sec:cont}
Continuum images, with a typical sensitivity of 0.12 mJy beam$^{-1}$, are used to calculate at 1.4 GHz star formation rate following \cite{Yun2001}, as
\begin{equation}
SFR_{1.4 GHz} = \frac{L_{1.4}(ergs ~s^{-1} ~Hz^{-1})}{1.7 \times 10^{28}} M_\odot ~year^{-1} .
\end{equation}
Continuum emission is detected in only four void galaxies, and only two are extended enough (VGS\_07 and VGS\_32) to rule out confusion with possible active galactic nucleus (AGN) emission, however all radio luminosities are low enough that the net AGN contribution should be small. For those galaxies without emission we calculate a 3$\sigma$ upper limit (Table~\ref{tab:starmass}), which we find to be in general agreement with the H$\alpha$ star formation rate (described below).

\subsection{Optical Data}
\label{sec:optical}
To determine  the stellar mass of  the galaxies from  the SDSS optical
spectra,  each spectrum is  pre-corrected for  dust extinction  due to the
Milky  Way  using  the  map  by  \citet{Schlegel1998}.   Adopting  the
approach used  by various authors \citep[e.g.,][]{Gwyn2005,Salim2007},
the stellar mass  is determined by fitting to  the full galaxy spectra
(3800--9200~\AA  \   in  the  observed   frame)  from  the   SDSS  DR6
\citep{sdssdr6}   the   dust-attenuated  \citet{Bruzual2003}   stellar
population model spectra, where the dust model of \citet{Calzetti2000} 
is  used to correct for intrinsic  dust extinction.  The
overall model  is composed of simple stellar  populations with stellar
age ranges  from 5~Myr to  11~Gyr and stellar metallicity  ranges from
0.004  to 0.05.   The dust  extinction is  a free  parameter  which is
determined simultaneously through the best-fit stellar continuum.  
Using a  galaxy sample that is all galaxies
included in both
the SDSS DR4 and the SDSS DR6 data,
the
agreement between our dust-free  stellar mass estimates and those from
the  line-indices approach  performed  upon the  SDSS DR4
galaxies \citep{Kauffmann2003} is on average $\sim${0.25}~dex (Yip et
al.  2010 in prep.).

As  the SDSS  spectroscopy samples  the light  of each  galaxy  in the
central  3$^{\prime\prime}$-diameter area,  an aperture  correction is
applied  in  deriving the  stellar  mass  of  the whole  galaxy.   The
correction  is taken  to  be the  ratio  between the  flux within  the
Petrosian radius \citep[which encompasses  more than $\sim${90}\% of a
galaxy regardless of the light profile,][]{Strauss2002} to that within
the  central 3$^{\prime\prime}$  \ diameter  of each  galaxy,  as such
$\mbox{m}_{*}^{\mathit{petro}}   =   \mbox{m}_{*}^{\mathit{fiber}}  \,
10^{-0.4   \,  (m^{\mathit{petro}}   -   m^{\mathit{fiber}})}$,  where
$\mbox{m}_{*}$ and $m$ denote the  stellar mass and the observed frame
magnitude   of  the  galaxy,   respectively.   
Stellar  masses  and
H$\alpha$  star  formation  rates,  as  determined  by  the  H$\alpha$
luminosity of the optical spectra with an aperture correction applied,
are given in Table~\ref{tab:starmass}.

Even though we are considering a galaxy population which may be systematically different, our stellar mass estimates are accurate within the tolerances of the method, and are relatively insensitive to the assumed parameters  for stellar age and metallicity.
Our  void galaxy  sample shows  an  agreement of $\sim${0.25}~dex between  our stellar mass estimates and  those in the MPA-JHU DR7  catalog \footnote[1]{The MPA-JHU catalog is publicly      available     and      may     be      downloaded     at http://www.mpa-garching.mpg.de/SDSS/DR7/}  which is based on fits to the photometry through an approach similar to \citet{Kauffmann2003}  and   \citet{Salim2007}.  
Our
choice of the modeled stellar age lower bound, 5~Myr, ensures that the
massive stars ($\sim{100}$~$M_{\odot}$) are included in the model.  On
the choice  of the modeled  stellar metallicities: the  stellar masses
are  expected  to  be   fairly  insensitive  to  the  modeled  stellar
metallicities,  because they  are derived  from the  stellar continuum
luminosity, whereas  the stellar  metallicity of the  galaxies affects
primarily  the absorption line  strengths.  On the other  hand, the
stellar  mix does determine  the amplitude  of the  underlying stellar
absorption  in  the  H$\alpha$  line.   The  average  H$\alpha$  stellar
absorption    in    the     SDSS    galaxies    is    ${1.3}$~\AA    \
\citep{Hopkins2003,Miller2003}  determined through  an  approach
which  is  independent  of  stellar  population model,  and  is  small
compared with the average  restframe H$\alpha$ emission equivalent width of our
galaxies, calculated to be about ${50}$~\AA.

\subsection{Notes on Individual Void Galaxies}
\label{sec:individual}
Many of our void galaxies appear morphologically and dynamically quite unexceptional.  VGS\_09, VGS\_14, VGS\_36, VGS\_42, VGS\_45, and VGS\_58 all appear to exhibit regular disk-like rotation in their H \textsc{i} contours, and a fairly smooth, disk-like stellar component.  At 3$\sigma$ contour levels their H \textsc{i} appears uniform and axisymmetric within resolution limits.  However, our small sample does include a number of notable individual galaxies.

\textbf{VGS\_07} has a faint companion which does not appear to be interacting. It exhibits a clumpy, knotty, blue morphology similar to that seen in chain-galaxies forming at high redshift, or like edge-on, low surface-brightness galaxies.  It has an enormous 200 \AA~H$\alpha$ equivalent width, suggesting a high star formation rate per unit stellar mass, and possible starburst.  We see fairly regular disk-like kinematics within the gas distribution.  It sits near the center of the void and not near a wall-like configuration.

\textbf{VGS\_12} is a polar ring galaxy, and is discussed in detail in \cite{Stanonik2009}.  It is one of the most massive of our sample, in both its H \textsc{i} ($3.0 \pm 0.5 \times 10^9 \; M_{\odot}$) and dynamical ($1.5 \times 10^{11} M_{\odot}$) mass, and is substantially more massive in H \textsc{i} than stellar mass ($1.05 \times 10^9 M_{\odot}$).
In addition, its H \textsc{i} disk is extremely extended compared to the optical, yet free of stars down to the SDSS surface brightness limits, ruling out most merger scenarios.
Simulations can reproduce such gas-rich, star-poor rings through tidal encounters, however at equal mass ratios they destroy the rotational support of the central galaxy, resulting in the formation of an elliptical remnant \citep{Bournaud2005}.
The low concentration index, $r_{90} / r_{50}$, of 2.25 implies the central stellar disk is late type \citep{Shen2003} and thus rotationally supported, suggesting an alternative formation mechanism such as slow accretion of cold gas is necessary \citep{Iodice2006}.  
In addition, VGS\_12 is situated in a thin wall between two voids, and aligned such that this gas would have been accreted out of the voids. 

\textbf{VGS\_30} is accompanied by a similar sized H \textsc{i} rich companion.  The two are seen to be embedded in a common H \textsc{i} envelope, and due to the thinness of the target galaxy disk, presumably they are interacting for the first time. They are cosmologically situated in a void wall between two large voids.

\textbf{VGS\_32} is particularly close by, and thus significantly better resolved in optical and H \textsc{i} imaging.  It displays a quite beautiful flocculent spiral pattern, and a textbook flattened rotation curve.  It sits nicely isolated in the middle of a void.

\textbf{VGS\_34} has a very faint, kinematically distinct protuberance in H \textsc{i}. The disturbed optical morphology also suggests that the target is undergoing some sort of interaction, as the bulge appears extremely red while the farther reaches of the disk appear relatively blue and distorted.  The H \textsc{i} extention does align with a small, distant H \textsc{i} detected companion.  VGS\_34 also has a very large rotational velocity.  Like VGS\_07, it sits near the center of the void and not near a wall-like configuration. Because the SDSS fiber only contains light from the central red bulge, stellar mass and star formation rate parameters derived for this galaxy may not accurately represent the galaxy as a whole.

\textbf{VGS\_35} presents a fairly regular appearance in H \textsc{i}, except in its orientation.  We see it exhibiting a strong warp, or perhaps a mis-aligned H \textsc{i} disk as we see in VGS\_12.  This is complicated by the elongated beam, which is suspiciously aligned with the H \textsc{i} in the disk.  Possibly the elongation of the beam is exaggerating a slight warp in the H \textsc{i} disk, as the velocity contours in the center appear aligned with the optical disk, and bend away only in the outer parts.  It is located near the transition of a small void into a larger one.

\textbf{VGS\_38} has the most optically disturbed appearance, and it was also the only selected void galaxy to have a known companion,  VGS\_38a, which is only 12 kpc away on the sky and at nearly overlapping velocity.  The surprising discovery of a third galaxy, VGS\_38b, aligned with the first two, all three of which are connected by a faint 1$\sigma$  H \textsc{i} bridge, results in a system that stretches nearly 50 kpc across the sky.  VGS\_38, despite its chaotic appearance, exhibits relatively smooth rotation in the direction perpendicular to the line of galaxies.  It is difficult to understand how this system could have undergone a major, disruptive interaction while leaving the gas disk unperturbed.  It is located at the edge of its void.

\subsection{Control Sample}
\label{sec:control}
\begin{figure}
\centering
\includegraphics[scale=1]{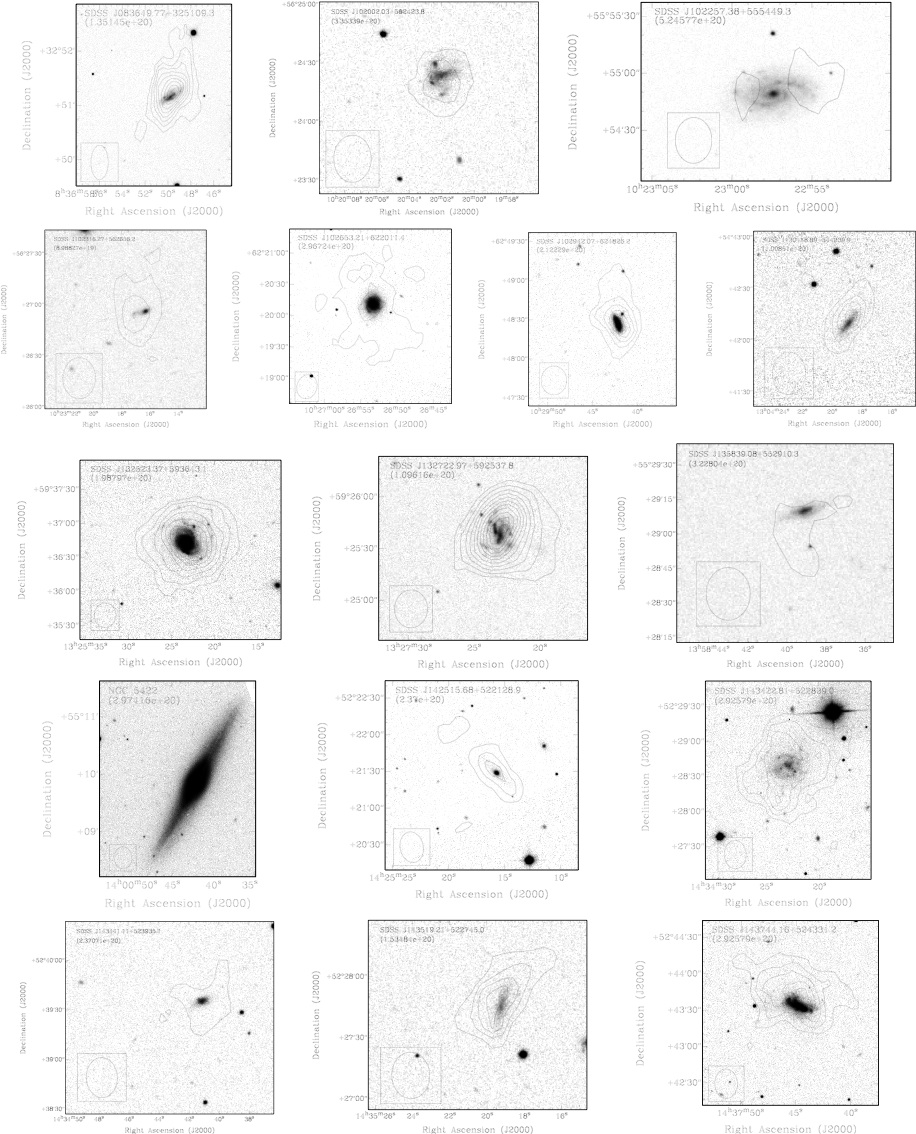}
\caption{Control sample.  Contours in the total intensity maps are at intervals of 2$\sigma$, with the corresponding column density value (in units of $cm^{-2}$) given in the top left corner of each image.  Note that low signal to noise results in relatively high 2$\sigma$ column densities of $\sim2 \times 10^{20} cm^{-2}$.  Contours for NGC 5422 represent a 1$\sigma$ detection.
\label{fig:ctrl}}
\end{figure}

\begin{figure}
\centering
\includegraphics[width=0.99\textwidth]{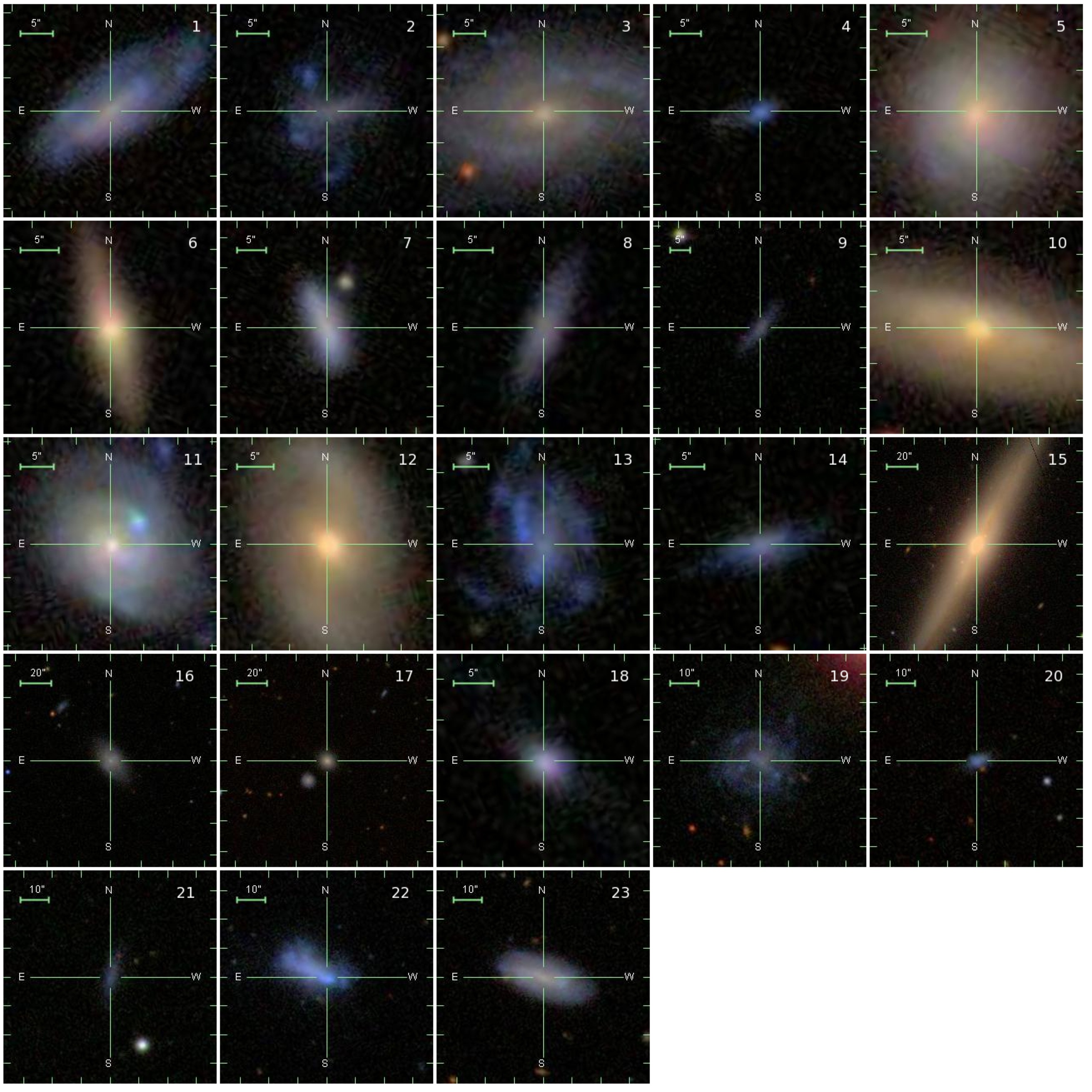}
\caption{Our control sample galaxies, scaled to the same physical size as the void galaxy sample in Figure \ref{fig:poststamps}. Composite color images are taken from the online SDSS Finding Chart tool. } 
\label{fig:poststampsctrl}
\end{figure}

From projection of each observation's primary beam through the SDSS redshift catalog we find that there are 22 SDSS galaxies in the neighboring velocity ranges of our data set.
From our H \textsc{i} blind  search for control sample galaxies (described in Section~\ref{sec:obs}), we detect 11 of the known SDSS galaxies and have upper limits for another seven, with mass limit $M_{\textrm{H \textsc{I}}}=2.9\times10^4$ d$^2$ $M_\odot$, d in Mpc and assuming a velocity width of 100 km s$^{-1}$.
The data on the remaining four galaxies were badly affected by 
man made interference and could not be used. 
We also include with this control sample five galaxies detected in H \textsc{i} which are not in the SDSS redshift survey, all with optical counterparts. 
This brings the control sample to a total of 23 galaxies.  
We calculate the environmentally filtered density contrast (see Section \ref{sec:select}) for each control sample galaxy, and find  almost all of them are in average or overdense environments.
The $g$-band images with H \textsc{i} contours of all H \textsc{i} detected control galaxies are presented in Figure~\ref{fig:ctrl}, and color images of all control galaxies are presented in Figure~\ref{fig:poststampsctrl}.  

To ensure confidence in our detections we only consider galaxies within 25$^\prime$ of the beam center, which corresponds to a factor of five reduced sensitivity.  The combination of larger distances and displacement from the beam center produces fairly low S/N detections,
resulting in larger mass uncertainties
than for our targeted void galaxies. The lowest H \textsc{i} contour levels for our control sample (Figure~\ref{fig:ctrl}) correspond to roughly twice the column density as the H \textsc{i} contours for our void galaxies (Figure~\ref{fig:vgs}).
The H \textsc{i} mass detection of NGC~5422 is actually less than the estimated error, however kinematics and spatial coincidence with the optical galaxy allow us to identify this as a real detection.  Parameters for all detected and nondetected control galaxies are listed in Tables~\ref{tab:altifs1}, \ref{tab:altifs2} and \ref{tab:starmass2}.

\section{Analysis} 
\label{sec:analysis}

\begin{figure}
\centering
\includegraphics[width=3.5in,angle=90]{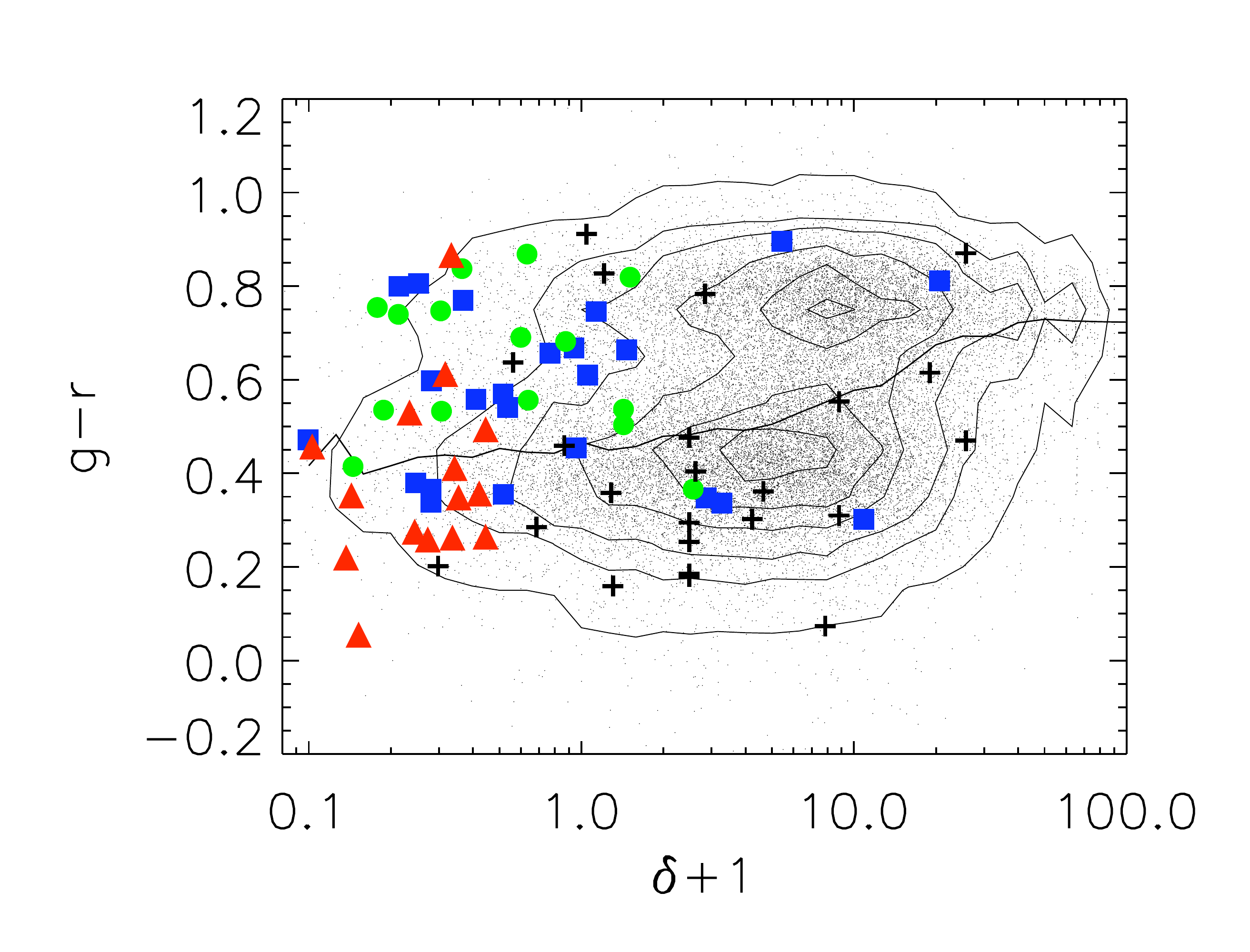}
\includegraphics[width=3.5in,angle=90]{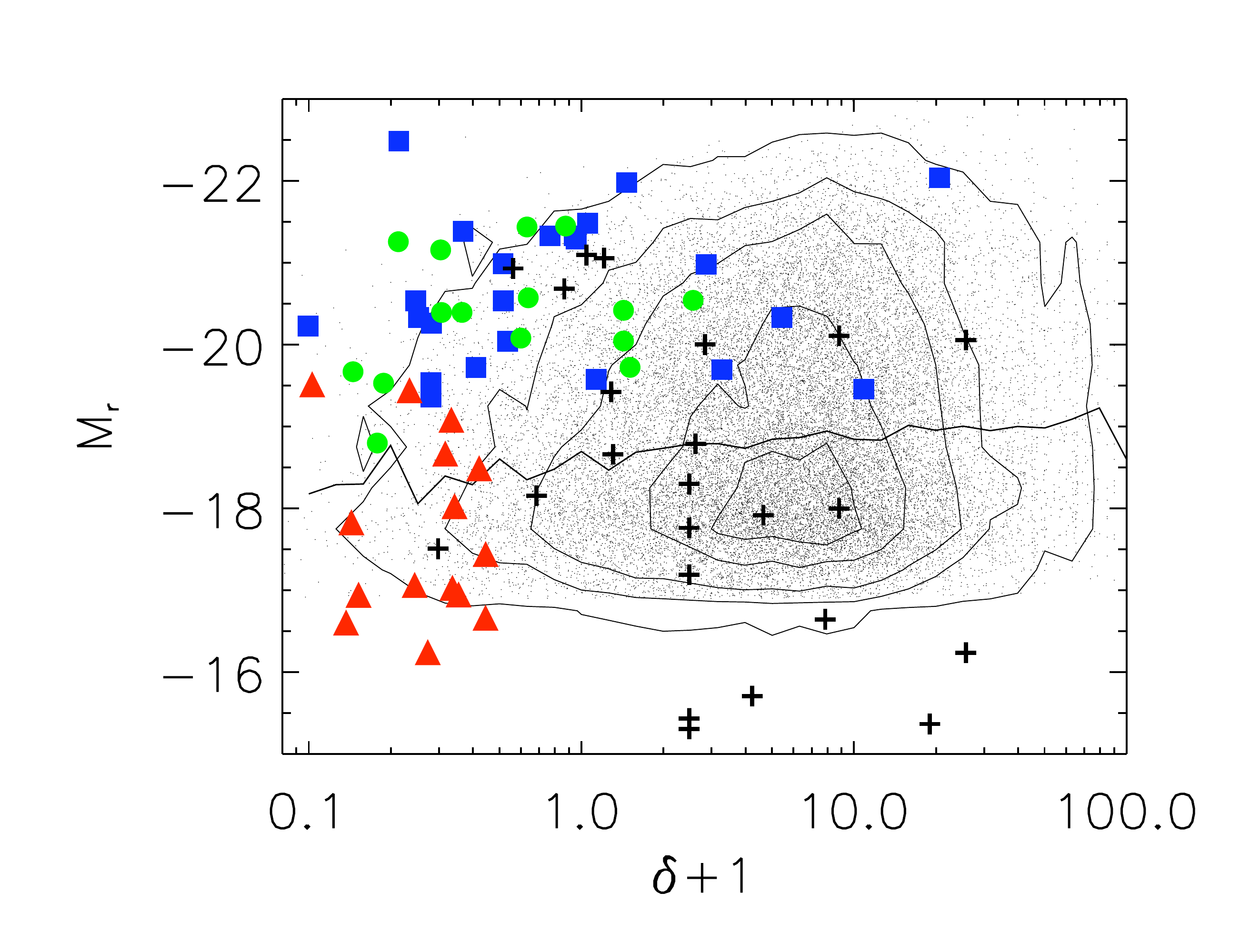}
\caption{ Distribution of $g-r$ colors and $r$-band absolute magnitudes for our void galaxy sample (triangles), our control galaxy sample (crosses), the Bo\"{o}tes void galaxy sample of \cite{Szomoru1996} (squares) and the CfA void galaxy sample of \cite{Grogin1999} (circles) as a function of density, compared with a volume limited ($z<0.02$, M$_r<-16.9$) sample of SDSS galaxies (points, with contours to guide the eye).  The line represents the median of the SDSS sample. 
\label{fig:dcm}  }
\end{figure}

Our void galaxy sample is carefully selected to represent the galaxies populating the deepest underdensities (Figure~\ref{fig:voidgalspatial3}, \ref{fig:voidgalspatial14}).  
For comparison we constructed a volume-limited ($z < 0.02$, $M_r < -16.9$) comparison sample of SDSS galaxies, and use our DTFE procedure to calculate the density field (see Section \ref{sec:select}).  
Eleven of our 15 void galaxies fall within these volume-limited criteria. 
In Figure~\ref{fig:dcm}, we compare the properties of the volume-limited sample as a function of density with our void sample, the  \textit{IRAS} selected Bo\"{o}tes void galaxy sample of \cite{Szomoru1996}, and the second Center for Astrophysics redshift survey (CfA2) lowest density void galaxy subsample of \cite{Grogin1999}.  Due to the incompleteness of past redshift surveys and brighter magnitude limits, the densities of the Bo\"{o}tes void galaxies and the CfA2 void galaxies have been recalculated based on their inclusion in the SDSS following our method described in Section~\ref{sec:select}, which results in the identification of many of these galaxies as being in average or even overdense environments.   
In 
this respect we emphasize the advantage of using the localized and better defined density values of the DTFE method, defined 
by the spatial galaxy distribution itself, as opposed to more conventional artificially filtered density values. 
As a result, our geometric selection criterion manages to probe a galaxy population which is truely representative of the extremely 
underdense and desolate void interiors, instead of the more conventional techniques probing the galaxies near the void boundaries.

The main photometric and spectroscopic comparison sample of  void galaxies against which we measure our pilot sample of geometrically selected void galaxies is that used by  \cite{Rojas2004} and \cite{Rojas2005}.  Their sample of 1010 void galaxies out to z=0.089 was compiled from the first SDSS data release, where they define a void galaxy as having the third nearest neighbor be at a distance greater than 10 Mpc.  All other galaxies are placed into a wall galaxy sample. Within these samples they constructed a bright (M$_r\le-17$) and nearby (d $<$ 103 Mpc) subsample of 76 void galaxies and 1071 wall galaxies.  Although our geometric method is very different, ten of our void galaxies satisfy these luminosity and distance criteria.

As has been noted by e.g. \cite{Rojas2004}, more luminous galaxies prefer denser environments, and our void galaxy sample is able to probe specifically those low luminosity galaxies which make up the bulk of the void galaxy population that were previously inaccessible.  Also apparent in Figure~\ref{fig:dcm} is the dominance of blue galaxies at the deepest underdensities, where we again are more accurately sampling the void galaxy population.
For our 10 bright, nearby void galaxies, we find a median $r$-band luminosity of -18.3 $\pm$ 0.3, and median $g-r$ color of 0.46 $\pm$ 0.06, which is in good agreement with the bright, nearby void galaxy sample of \cite{Rojas2004}.  We also note that our control sample is fairly representative of the range of absolute magnitudes, colors, and densities contained within the SDSS.

\begin{figure}
\centering
\includegraphics[height=4in,angle=90]{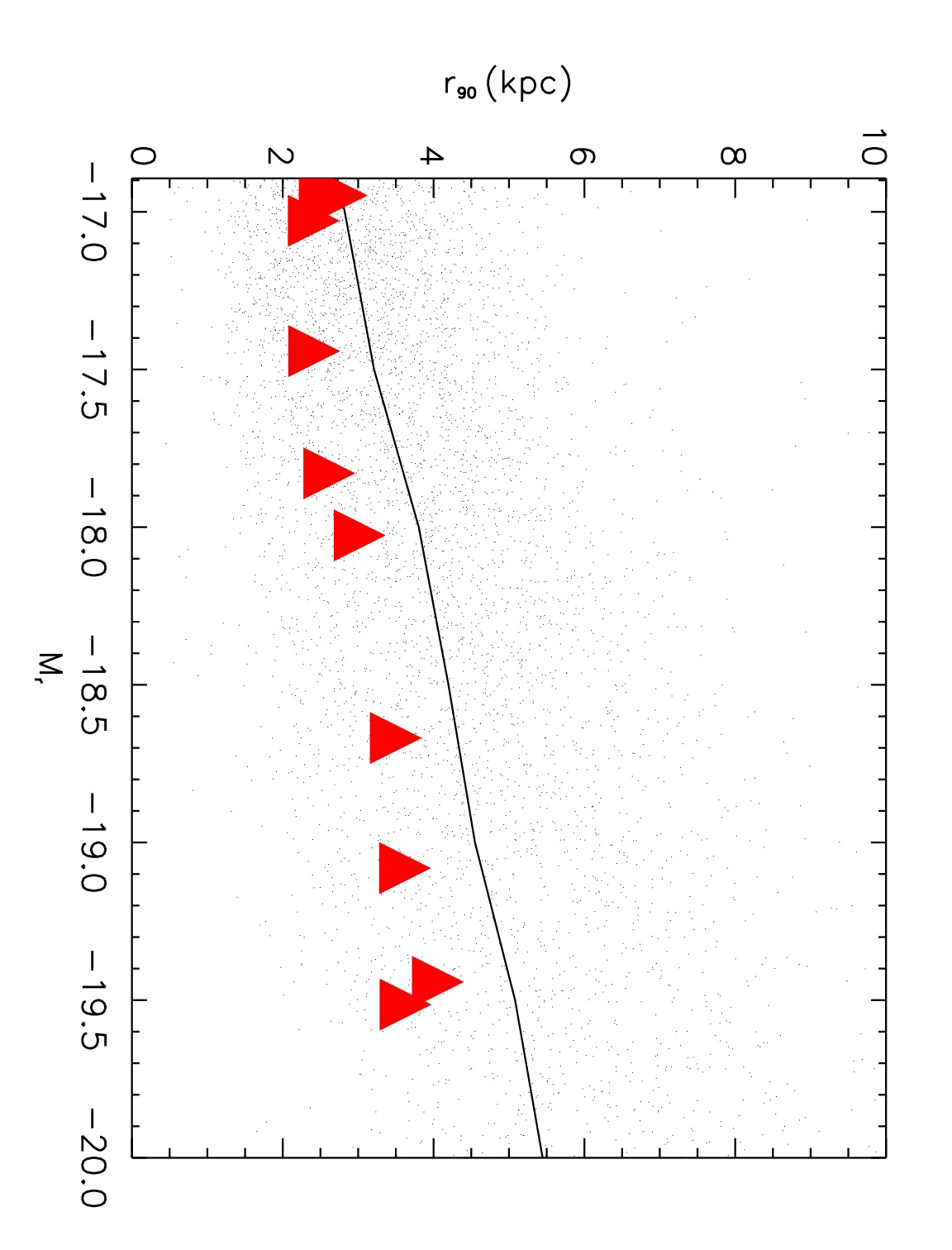}
\caption{The $r$-band r$_{90}$ radii of the stellar disk.  Our late-type void galaxies (triangles) fall systematically below the median (line) of a volume limited SDSS sample of late-type galaxies (dots). 
\label{fig:mrr90}}
\end{figure}

While we have selected void galaxies with a representative distribution of absolute magnitudes, we find that our void galaxies have stellar disks that are smaller than average.  Compared with the volume limited SDSS sample, we find the $r$-band r$_{90}$ radius of our late type void galaxies is systematically lower than the median for late type galaxies (Figure~\ref{fig:mrr90}), where galaxies with a concentration index r$_{90}$/r$_{50}$ $<$ 2.86 are taken to be late type following \cite{Shen2003}. 

While our targeted void galaxies are small, they would generally not be classified as dwarf galaxies.  All have M$_r <$ -16 and exhibit moderate circular velocities, typically 50-100 km s$^{-1}$. All galaxies exhibit signs of rotation in H \textsc{i}, though limiting resolution and lower sensitivity at the disk outskirts means we do not always see a flattening of the rotation curve.  $M_{dyn}$ is typically 10$^9$ to 10$^{10} M_\odot$.  The companion galaxies detected are mostly dwarf galaxies, however our small number statistics for this pilot sample limit us from extending these few detections to draw conclusions on the missing dwarf galaxy population pointed out by \cite{Peebles2001}.  

\begin{figure}
\centering
\includegraphics[width=2.3in,angle=90]{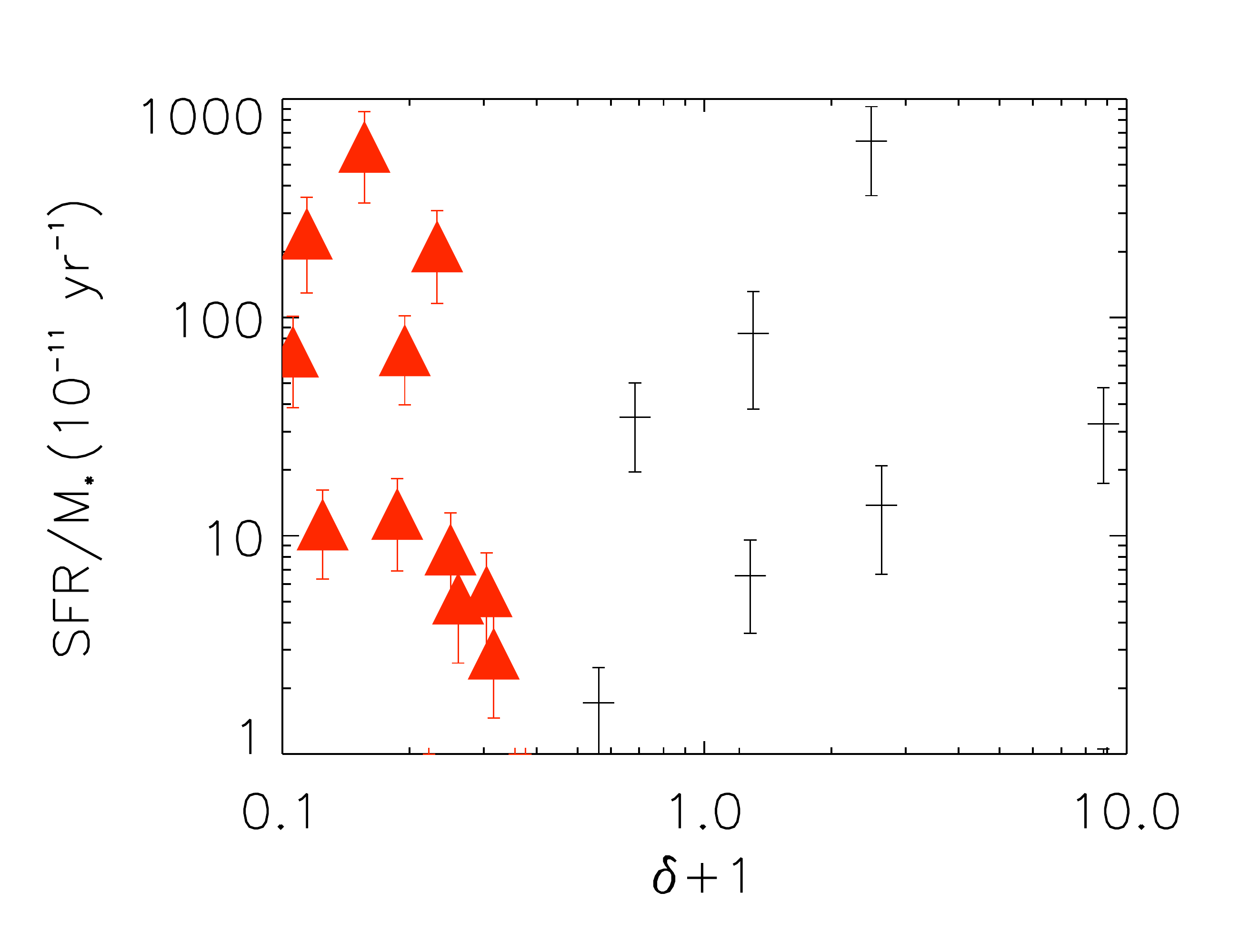}
\includegraphics[width=2.3in,angle=90]{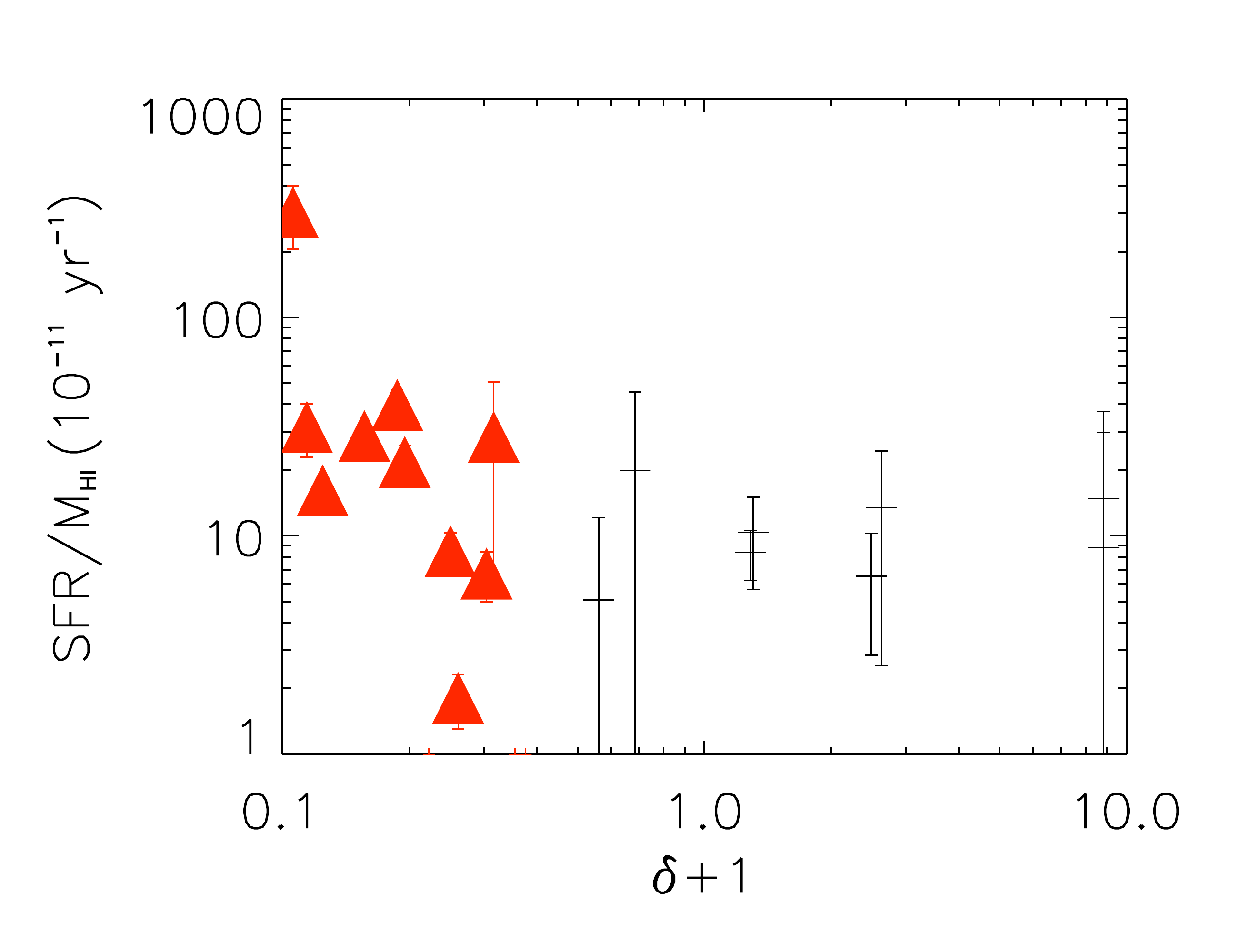}
\caption{S-SFR (left) and SFR/H \textsc{i} mass (right) as a function of density for our void galaxies (triangles) and our control sample (crosses).
\label{fig:sfrhi}}
\end{figure}

The stellar and star formation properties of our nearby, bright sample measured from the SDSS spectra are in general agreement with the values found for the nearby, bright void galaxy sample of  \cite{Rojas2005}, and show our pilot sample to be a good representation of the actively star-forming galaxies typically found in voids (Table~\ref{tab:rojas}).   Even our small sample exhibits the same trend for increased H$\alpha$ equivalent width, star formation rate (SFR) and specific SFR (star formation rate per stellar mass, S-SFR) in low density environments when compared with the wall sample of \cite{Rojas2005}.  
Despite these average increases, we do not see a clear trend of S-SFR with density, but  we do see a hint of a trend for increased SFR per H \textsc{i} mass at the lowest densities (Figure~\ref{fig:sfrhi}).  The mean S-SFR for our full void sample is $(88 \pm 45) \times 10^{-11}$ year$^{-1}$ and for our control sample is $(82 \pm 63) \times 10^{-11}$ year$^{-1}$.  The mean SFR per H \textsc{i} mass is $(37 \pm 22) \times 10^{-11}$ year$^{-1}$ for our void sample and $(11 \pm 2) \times 10^{-11}$ year$^{-1}$ for our control sample. Our complete sample of 60 void galaxies will be more suitable for the identification of possible trends with density. 

Because our control sample for this pilot study is not very large, we have drawn from the literature two additional comparison samples with which to compare H \textsc{i} properties.  
The first is a sample of 113 late-type dwarf and irregular galaxies from the WHISP project (Westerbork observations of neutral Hydrogen in Irregular and SPiral galaxies, \citealt{vdHulst2001}) observed with the WSRT, selected to have Hubble type later than Sd or magnitude fainter than $M_B  = -17$ \citep[S02]{Swaters2002}.  
These galaxies were not selected in any way for environment, and in general their cosmological environments are unknown.
In an attempt to control for environment, we also 
compare with the H \textsc{i} imaging survey of 43 Ursa
Major galaxies  brighter than -16.8 and more inclined than 45$^\circ$ with the WSRT  \citep[VS01]{Verheijen2001}.
 Because of their shared membership in the Ursa Minor poor cluster, we assume some uniformity of environment within this comparison sample.

\begin{figure}
\centering
\includegraphics[height=4in,angle=90]{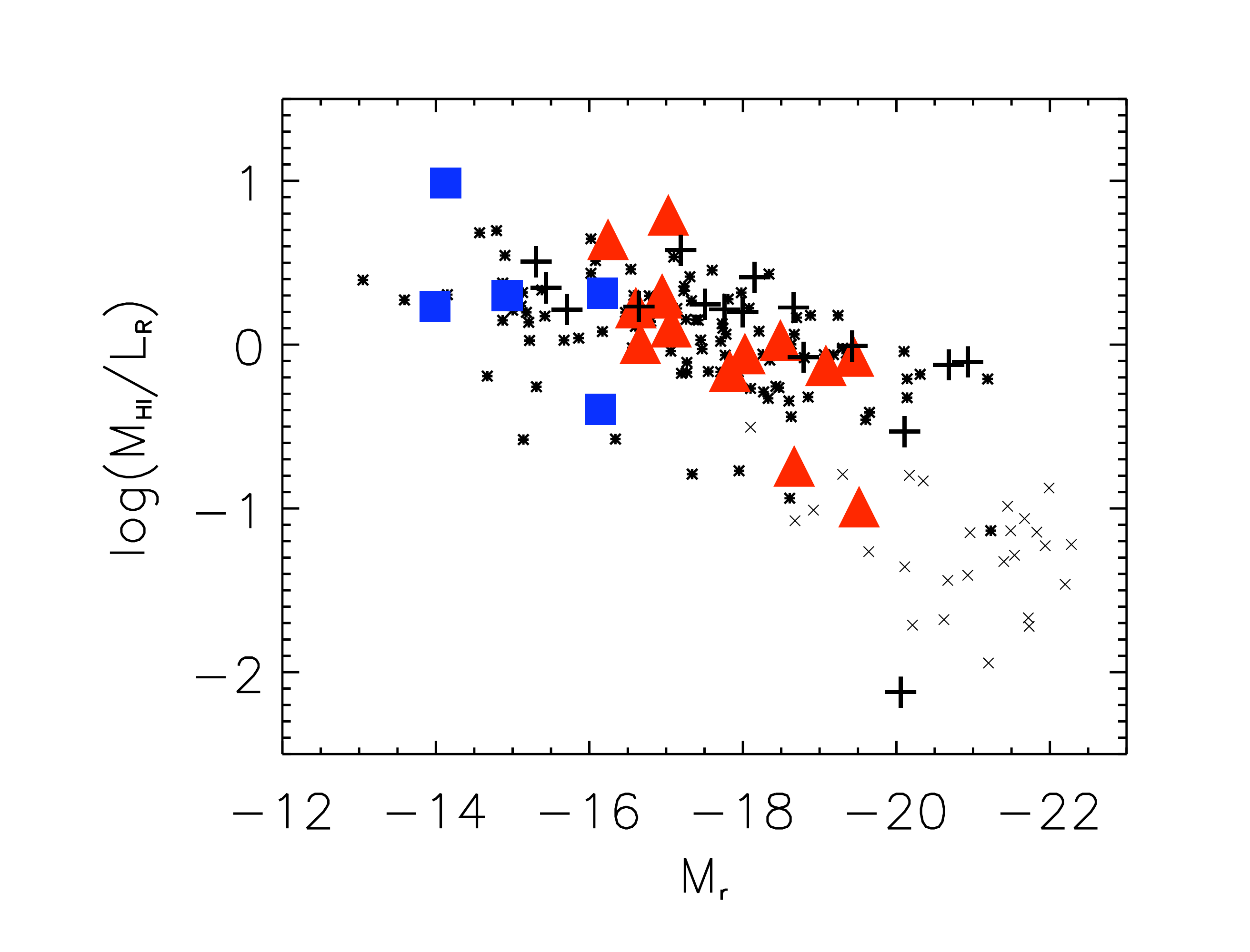}
\caption{H \textsc{i} mass to light ratio for our targeted void galaxies (triangles),  companion  galaxies (squares), and our control galaxies (crosses). For additional comparison,  late-type disk galaxies from \cite{Swaters2002} (stars) and Ursa Major galaxies from \cite{Verheijen2001} (X's) are included. 
\label{fig:plot2}}
\end{figure}

\begin{figure}
\centering
\includegraphics[height=4in,angle=90]{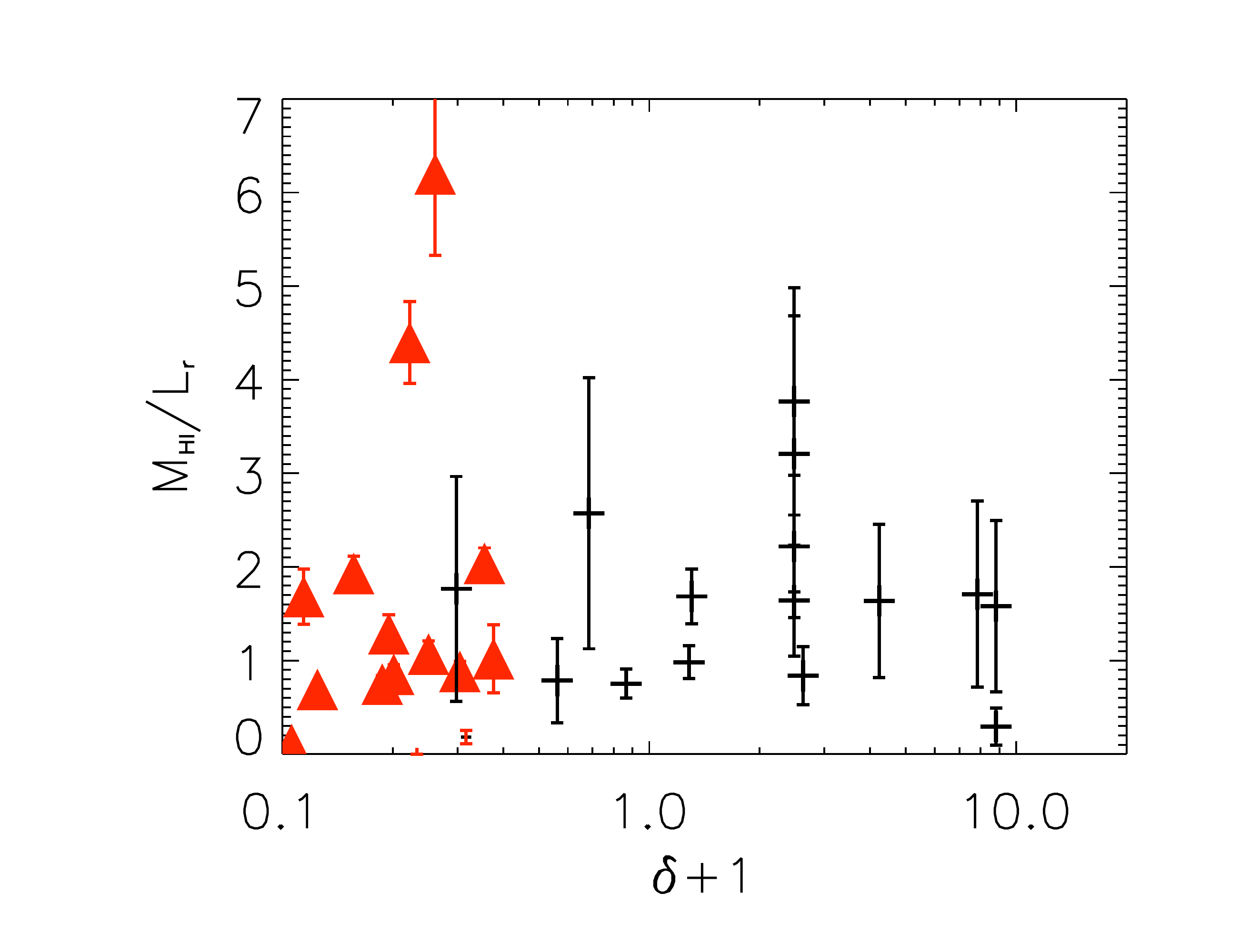}
\caption{H \textsc{i} mass to light ratio as a function of density for our void galaxies (triangles) and our control sample (crosses). 
\label{fig:mhilrdens}}
\end{figure}

While optically our sample is blue with higher star formation rates than galaxies in average density environments, the H \textsc{i} content appears to be more typical.  As in \cite{Geha2006}, we find that the smallest galaxies are the most gas-rich.  Figure~\ref{fig:plot2} shows that the ratio of $M_{\textrm{H \textsc{I}}}/L_r$ increases with decreasing luminosity similar to S02 and VS01.
This trend seems to persist over a wide range of luminosities and environments.  
\cite{Huchtmeier1997} measured that $M_{\textrm{H \textsc{I}}}/L_B$ increases with decreasing density. We do not see such a trend with density within our small sample of void galaxies (Figure~\ref{fig:mhilrdens}), nor do we see a significant difference between our void sample and control sample, or compared to the higher density comparison samples of S02 or VS01 (Figure~\ref{fig:plot2}).

\begin{figure}
\centering
\includegraphics[height=3.5in,angle=0]{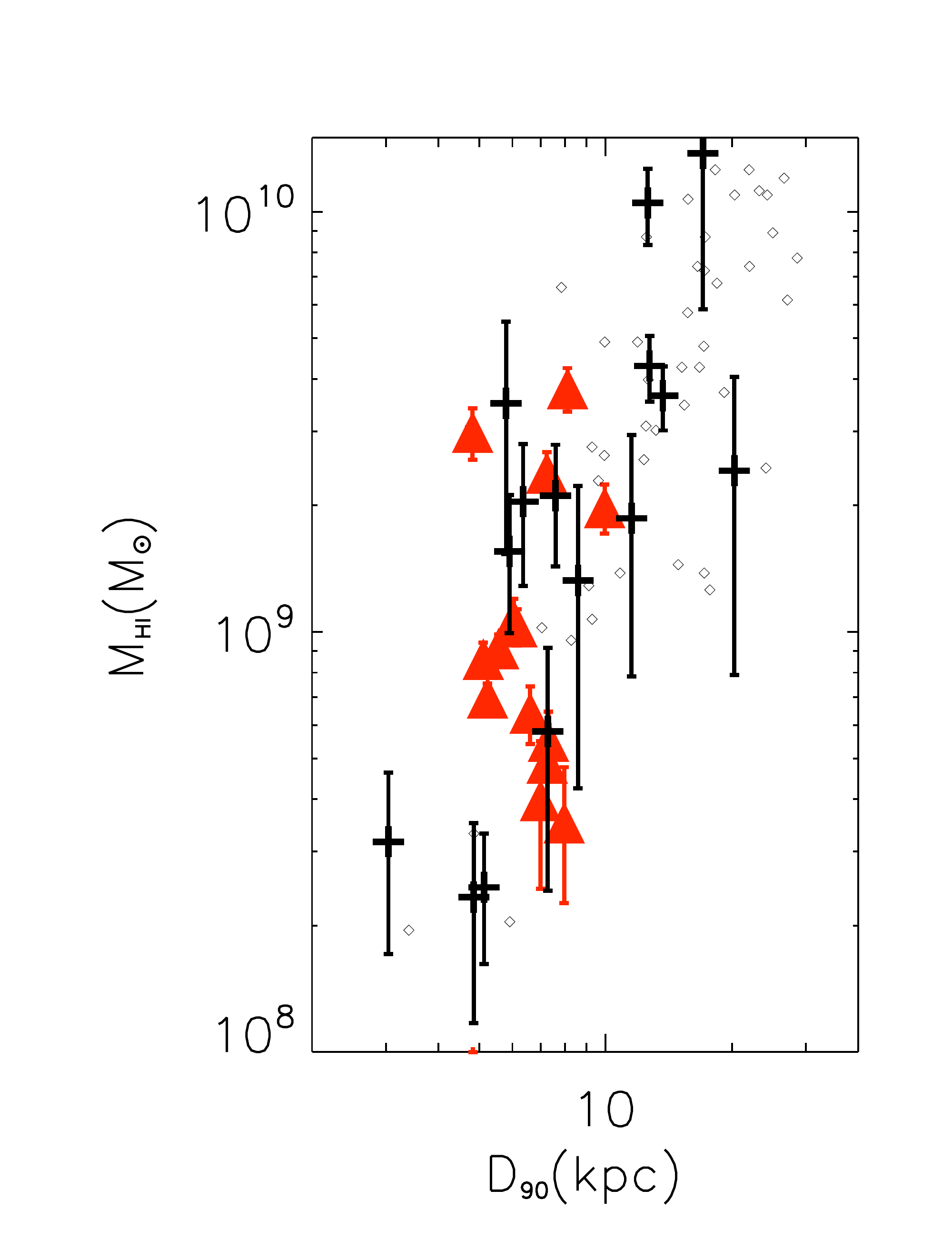}
\includegraphics[height=3.5in,angle=0]{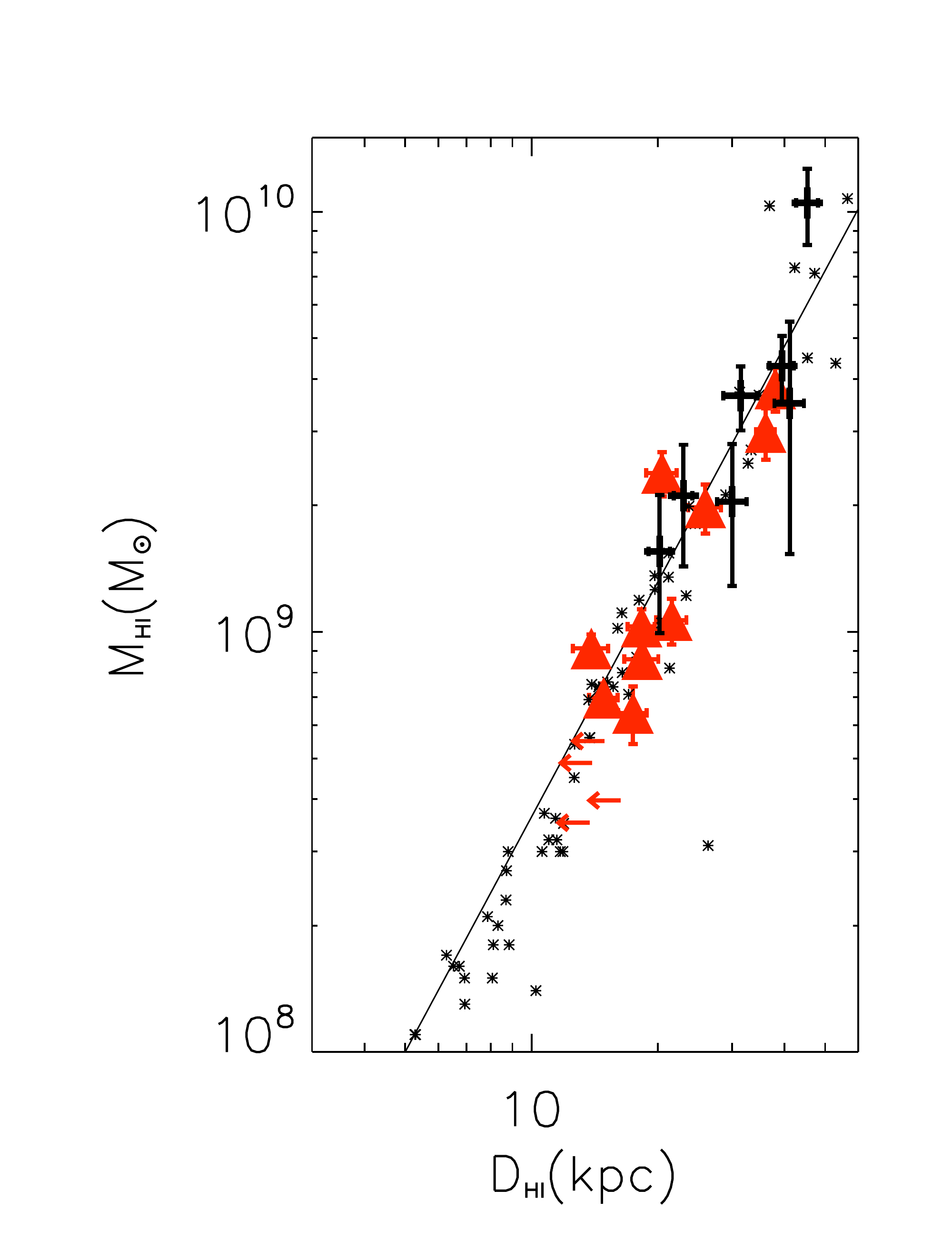}
\caption{H \textsc{i} mass versus optical diameter on the left, and versus H \textsc{i} diameter on the right. Our target galaxies (triangles) and control sample (crosses) are compared with the sample of late-type galaxies by \cite{Swaters2002} at right (stars) and a cross-match of ALFALFA galaxies \citep{Giovanelli2007} with the SDSS catalog at left (diamonds).  Overplotted on the right is the fit from  \cite{Verheijen2001} for their sample of Ursa Major galaxies. Red arrows mark the upper limit on the diameter for those void systems which are poorly resolved in H \textsc{i}.
\label{fig:massdiam}}
\end{figure}

Our void galaxies follow the tight relation between H \textsc{i} mass and H \textsc{i} diameter demonstrated by S02 and VS01 (Figure~\ref{fig:massdiam}, right).  The agreement of our underdense sample with both comparison samples shows that the mean H \textsc{i} surface density is quite robust, regardless of environment.  
Galaxy samples also show a correlation between H \textsc{i} mass and optical diameter (e.g. \cite{Verheijen2001}).  The range of optical diameters of our void galaxies is too small to look for a trend, but the measured values seem reasonable (Figure~\ref{fig:massdiam}, left).  Our control sample does appear to lie on the correlation.
We also note that while the resolution is not optimal, our H \textsc{i} disks typically appear roughly exponential in shape at their outer extent (Figure~\ref{fig:surfdens}), as was found by S02 for their late type disk galaxies.

\begin{figure}
\centering
\includegraphics[width=2.3in,angle=90]{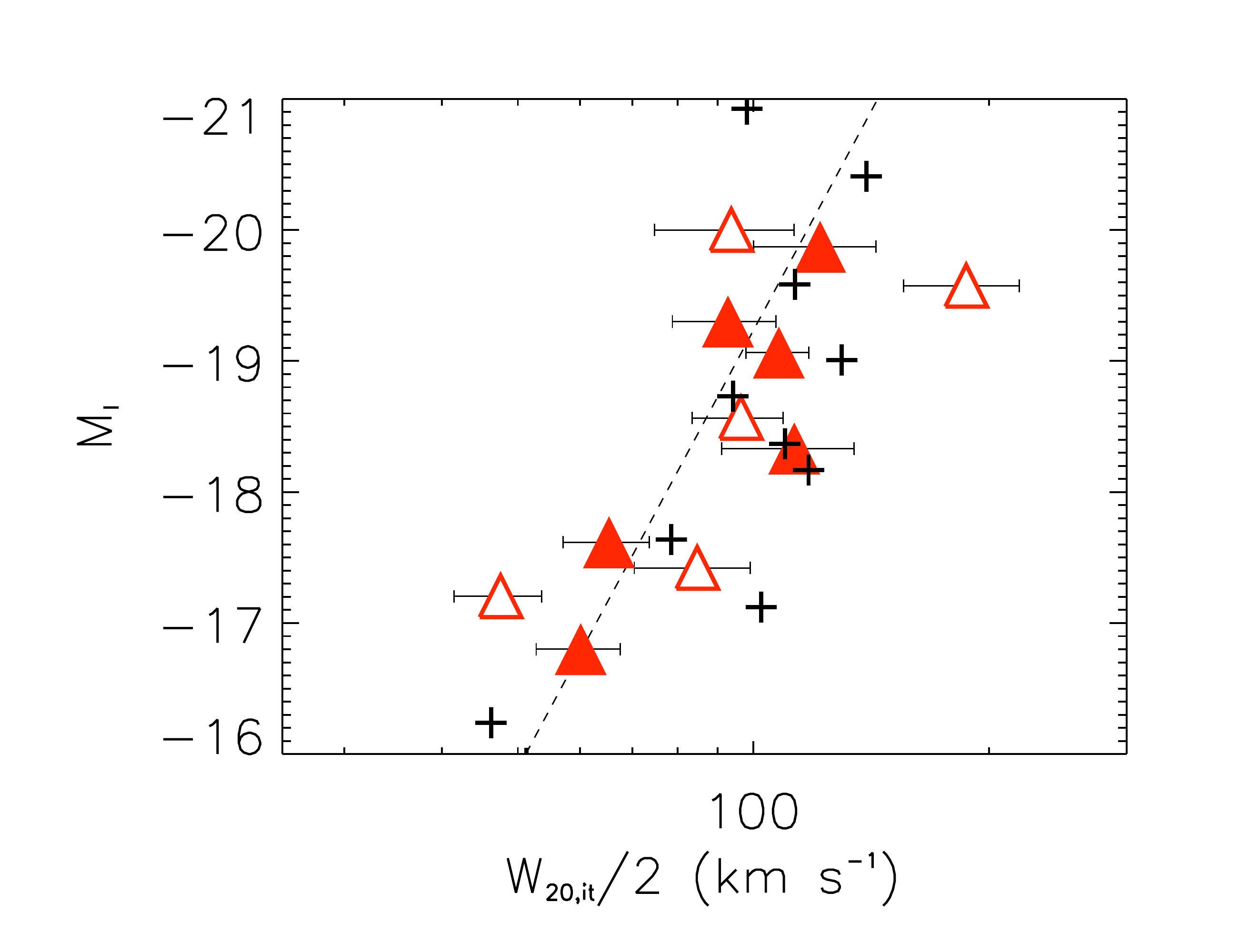}
\includegraphics[width=2.3in,angle=90]{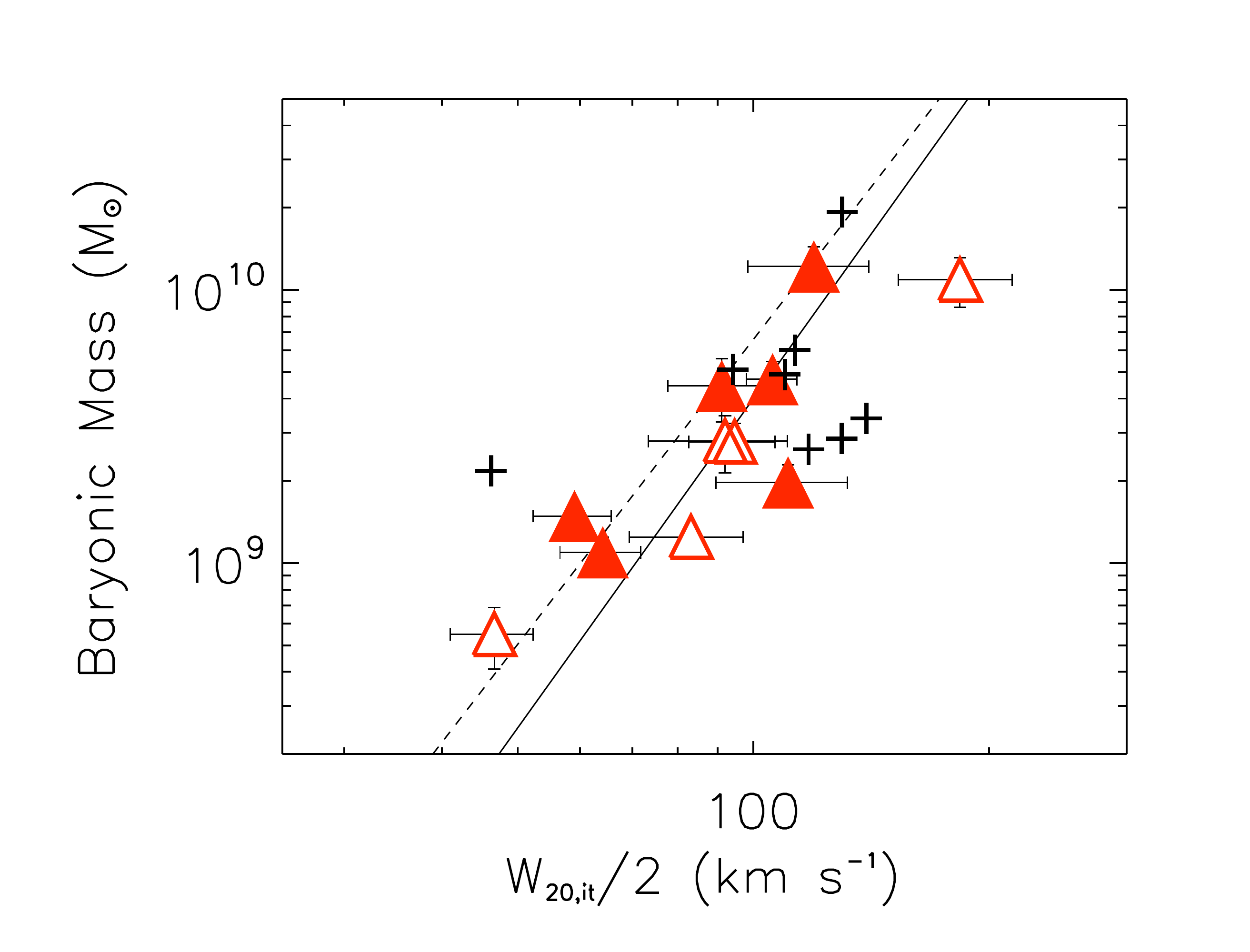}
\caption{Void galaxies following the I-band (left) and baryonic (right) Tully-Fisher relationship, where the filled symbols are galaxies for which we observe a flattened rotation curve and the open symbols are systems where the rotational velocity continues to rise at the furthest radii. Crosses indicate our control sample galaxies.  The dashed lines were taken from \cite{Geha2006} which were fit to over three orders of magnitude mass range. The solid line shows the fit taken from \cite{McGaugh2000}.    
\label{fig:tf}}
\end{figure}

\cite{Hoeft2006} predict the effects of UV background ionization will suppress star formation and lead to a reduced baryon fraction in small, $M_{tot}\sim10^9 M_\odot$ halos.  Our halos are just on the verge of this boundary, with $M_{dyn} > 2 \times 10^9 M_\odot$, however we see no deviation from the baryonic Tully-Fisher relationship (Figure~\ref{fig:tf}). W$_{20}$ values have been corrected for turbulent velocity broadening and inclination angle.  Error bars reflect a 10\% uncertainty in inclination.  Baryonic masses have been calculated from the sum of the stellar mass and 1.4 times the H \textsc{i} mass to estimate the total gas component from our H \textsc{i} measurement.  In some of our systems (open triangles) we do not observe a flattening of the rotation curve due to poor resolution or truncated H \textsc{i} disks.  As expected, better determined systems (filled triangles) show a tighter correlation and are more consistent in their scatter about the Tully-Fisher relationship, as derived in environmentally insensitive samples \citep{McGaugh2000,Geha2006}.

\section{Speculation}
\label{sec:speculation}
With such a small sample, it is difficult to discuss the statistical properties of our void galaxy sample. It is tempting, instead, to consider the situation of individual galaxies in our sample and the way they might illustrate the larger picture of galaxy evolution in voids which we are interested in investigating.

In some respects we expect to find galaxy evolution progressing within voids somewhat independent of their large scale environment. Simulations show that halos in underdense regions have assembled at later times \citep{Gao2005}, so we would expect that the galaxies we find there to simply be in a slightly earlier stage of their development.  This is reflected in the increased number of blue, star forming galaxies found in voids \citep{Rojas2004,Rojas2005}. However the formation of red elliptical galaxies by mergers is not excluded, and in fact the galaxies in the very deepest underdensities cluster more strongly than those in more moderately underdense regions, though less strongly than in average density regions \citep{Abbas2007}.  
This is similar to the conclusions of \cite{Szomoru1996}, who found a similar number of H \textsc{i} companions for void and cluster galaxies.  The clustering of dwarf companions appears somewhat consistent, regardless of environment \citep{Weinberg1991}.
Our sample probes to lower H \textsc{i} limits than \cite{Szomoru1996}, though they covered a much larger volume, and already we find a third of our void galaxies have previously unidentified companions, two of which share a common H \textsc{i} envelope with the targeted void galaxy.

We may expect environment will play some role. Cold mode accretion dominates in low mass halos and at earlier times \citep{Keres2009}, so we expect to see evidence of this ongoing slow accretion, though it may be difficult to distinguish from the effects of minor mergers. 
As void galaxies are statistically bluer at fixed luminosity than galaxies at higher densities \citep{Park2007,Bendabeck2008}, 
perhaps this is evidence of different mechanisms of gas accretion at play. VGS\_12, our void galaxy with a polar disk of H \textsc{i} gas,  might provide direct evidence that these processes do occur within voids, as it is difficult to amass such a substantial amount of H \textsc{i} without disrupting the rotational support of the central stellar disk \citep{Bournaud2005, Stanonik2009}.
We also note that the properties which do not seem to be sensitive to the low density environment ($M_{\textrm{H \textsc{I}}}/L_r$, the ratio of H \textsc{i} mass to H~\textsc{i} diameter, the Tully-Fisher relation) are more indicative of the internal effects within the galaxies.  The SFR, which we find is tentatively higher in voids, is instead indicative of an external effect, such as higher gas inflow.

Additional evidence of environmental effects should be evident. Hierarchical merging through the expansion of voids and thinning of void walls results in the formation of substructure within voids.  Filaments in dark matter which are quite distinct in simulations are tenuously traced by dark matter halos, but depending on the halo occupation distribution may not be apparent in the galaxy distribution, particularly when limited to the distribution of bright galaxies.  
These alignments may be more apparent in H \textsc{i} as with VGS\_38, where the linear alignment of all three galaxies, including the H \textsc{i} bridge between them, may already be sufficient to identify a filament within the void that is not apparent in the SDSS redshift survey. 

Filaments which are too low density to be traced by galaxies may still be traced by low column density H \textsc{i} or warm ionized gas \citep{McLin2002,Manning2002,Stocke2007}.  The intervening regions between the aligned components we detect near our void galaxies are ideal locations to use background quasars as probes and search for absorption due to underlying filamentary structure. 

\section{Conclusion}
\label{sec:conclusion}
We have completed a pilot study imaging in H \textsc{i} 15 void galaxies selected from the SDSS and located in d $<$ 100 Mpc voids.  Galaxies were selected purely by their environment to be located in areas less than half the cosmic density.  We detected 14 of our target galaxies with masses from $3.5 \times 10^8$ to $3.8 \times 10^9 M_\odot$, and had one nondetection with a 3$\sigma$ upper limit of $2.1 \times 10^8 M_\odot$.  In addition we detected six companion galaxies in H \textsc{i}, two of which appear to be interacting with the targeted void galaxy. 

We also constructed a control sample from the SDSS redshift survey galaxies and H \textsc{i} detected galaxies in the $\sim$10,000 km s$^{-1}$ velocity range simultaneously imaged by the WSRT behind and in front of the target galaxy with each pointing.  Of 18 detectable SDSS cataloged galaxies contained within the volume, we detected 11 in H \textsc{i} with masses $6.0 \times 10^7 M_\odot$ to $1.4 \times 10^{10} M_\odot$. In addition, we include five galaxies detected in H \textsc{i} which were not contained in the SDSS redshift catalog.  Because of their location within the SDSS footprint we were able to environmentally locate these systems as almost all being in regions at or above the cosmic mean.  Due to the small size of this control sample, when comparing specific H \textsc{i} properties of our void galaxies we also looked to comparison samples of well studied galaxies taken from the literature with similar size and luminosity constraints, though their environments are not known.

Provisionally, we find that our void galaxies have small optical stellar disks, typical H \textsc{i} masses for their luminosity, and generally follow the Tully-Fisher relation.  Consistent with previous surveys, they have increased rates of star formation, with the suggestion of a trend for increased star formation at lowest density.   
While this pilot sample is too small for any statistical findings, we did discover many of our targets to be individually interesting dynamically and kinematically in their H \textsc{i} distributions.  In particular, one shows possible evidence of ongoing cold mode accretion. 

Ultimately, we aim to compile a sample of $\sim$60 void galaxies which will comprise a new Void Galaxy Survey (VGS).  This sample size will be comparable to the H \textsc{i} sample of 24 void galaxies and 18 H \textsc{i} detected companion galaxies in Bo\"{o}tes by  \cite{Szomoru1996}, but more representative of the faint void galaxy population in a more distributed collection of voids. Our sample will also provide a good compliment to the sample of 76 SDSS void galaxies examined photometrically and spectroscopically by  \cite{Rojas2004} in similarly nearby voids, and will be ideal for a careful investigation of the questions surrounding how void galaxies get their gas, form substructures and generally populate the most underdense regions of the universe.

\acknowledgments
KS thanks Tony Wong for thought provoking discussion and advice.  RvdW is very grateful to Michael Vogeley, Bernard Jones and Ravi Sheth for the   discussions and ideas that inspired this project's initiative. We thank George Rhee and Tom Jarrett for their useful ideas and
  initiative in pursuing follow-up NIR observations. We also thank the anonymous referee for constructive comments which helped improve this paper.  
  This work was supported in part by the National Science Foundation under grant \#0607643 to Columbia University.
  We are grateful for support from a Da Vinci Professorship at the the Kapteyn Institute.
  We also acknowledge the Aspen Center for Physics and the Royal Netherlands Academy of
  Arts and Sciences (KNAW) for the generous and valuable support of the Aspen workshop
  on Cosmic Voids (June 2006) and the KNAW Colloquium on Cosmic Voids (December 2006),
  which were instrumental in defining our Void Galaxy Survey.

Funding for the SDSS and SDSS-II has been provided by the Alfred P. Sloan Foundation, the Participating Institutions, the National Science Foundation, the U.S. Department of Energy, the National Aeronautics and Space Administration, the Japanese Monbukagakusho, the Max Planck Society, and the Higher Education Funding Council for England. The SDSS Web Site is http://www.sdss.org/.

The SDSS is managed by the Astrophysical Research Consortium for the Participating Institutions. The Participating Institutions are the American Museum of Natural History, Astrophysical Institute Potsdam, University of Basel, University of Cambridge, Case Western Reserve University, University of Chicago, Drexel University, Fermilab, the Institute for Advanced Study, the Japan Participation Group, Johns Hopkins University, the Joint Institute for Nuclear Astrophysics, the Kavli Institute for Particle Astrophysics and Cosmology, the Korean Scientist Group, the Chinese Academy of Sciences (LAMOST), Los Alamos National Laboratory, the Max-Planck-Institute for Astronomy (MPIA), the Max-Planck-Institute for Astrophysics (MPA), New Mexico State University, Ohio State University, University of Pittsburgh, University of Portsmouth, Princeton University, the United States Naval Observatory, and the University of Washington.

\begin{figure}
\centering
\includegraphics[width=\textwidth]{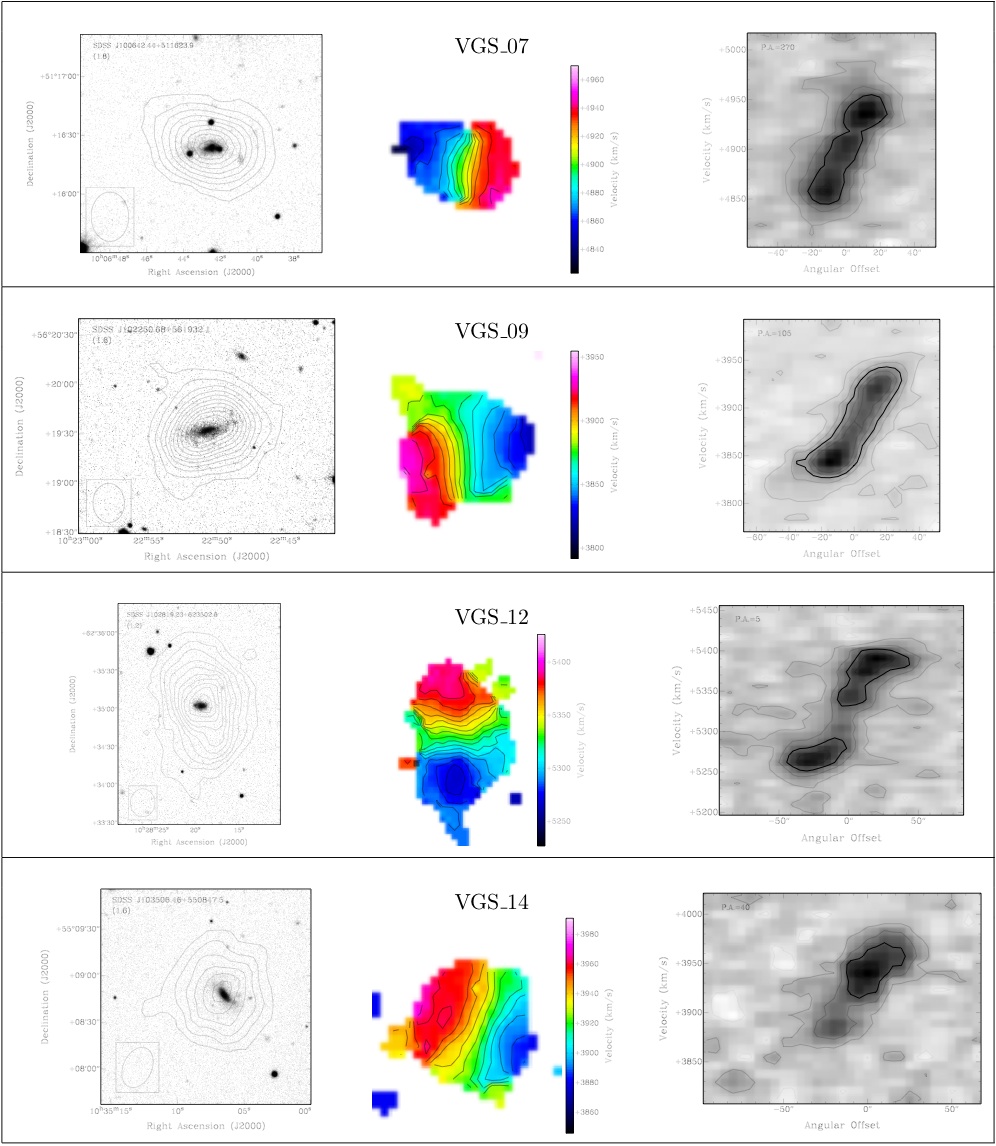}
\caption[]{}
\label{fig:voidgala}
\end{figure}

\begin{figure}
\centering
\ContinuedFloat
\includegraphics[width=\textwidth]{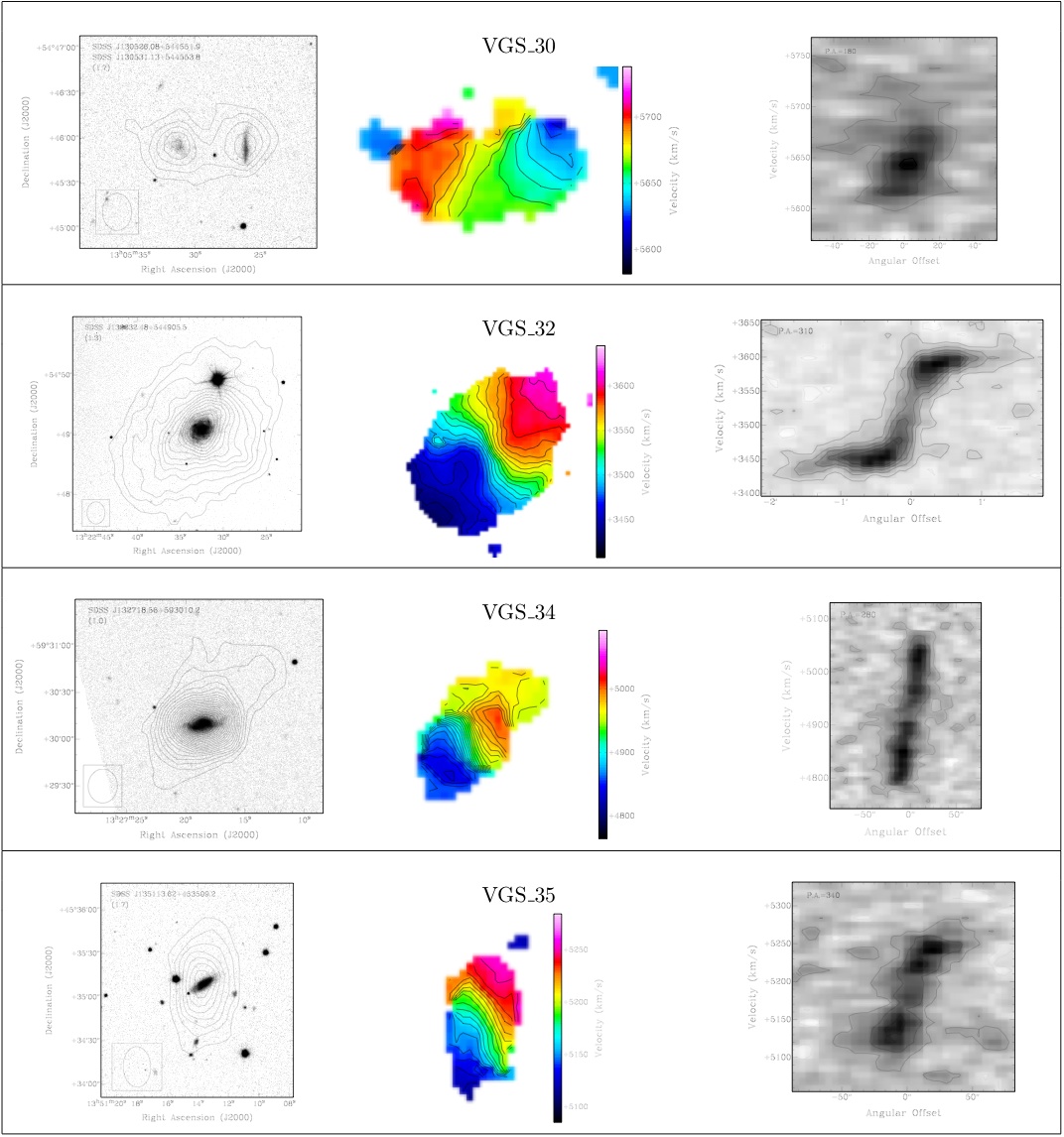}
\caption[]{}
\label{fig:voidgalb}
\end{figure}

\begin{figure}
\centering
\ContinuedFloat
\includegraphics[width=\textwidth]{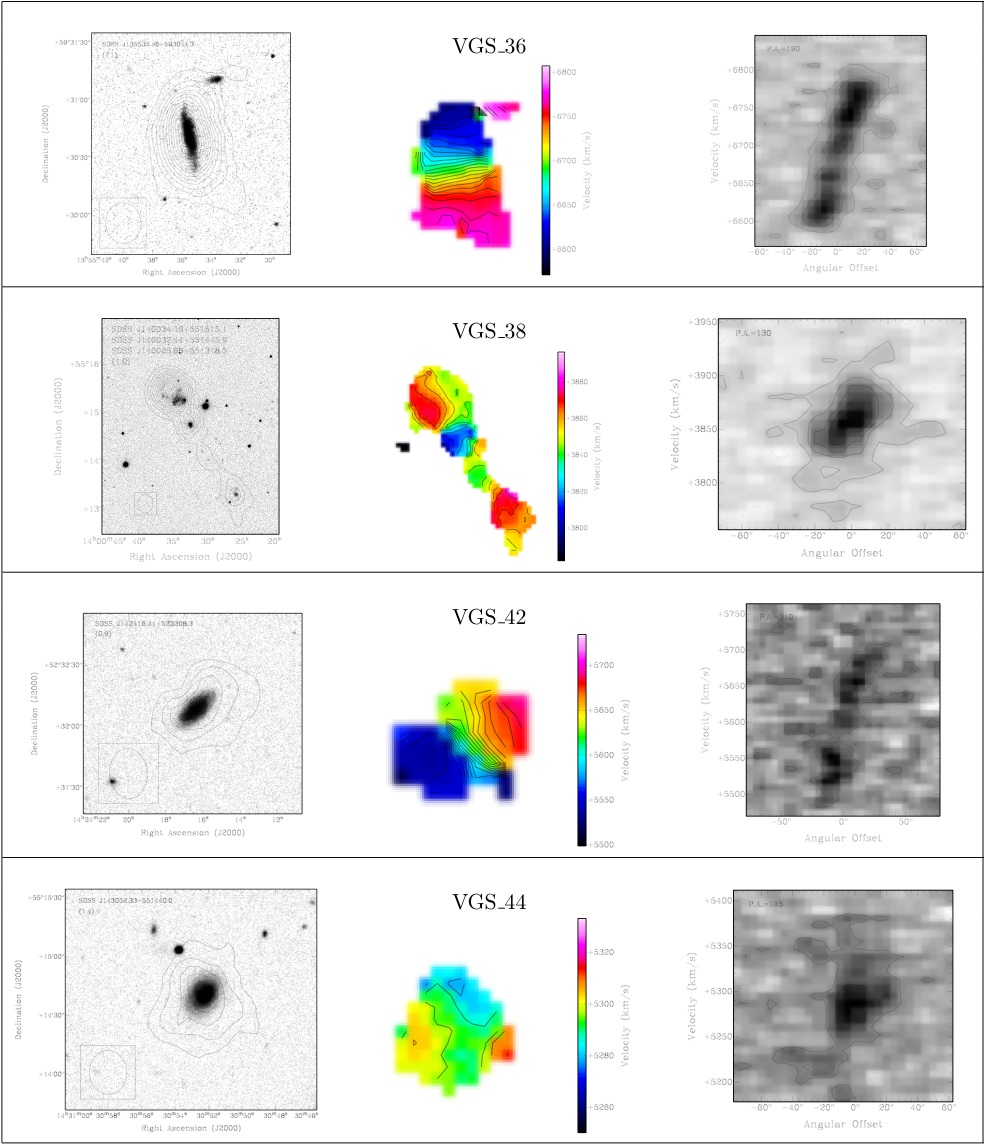}
\caption[]{}
\label{fig:voidgalc}
\end{figure}

\begin{figure}
\centering
\ContinuedFloat
\includegraphics[width=\textwidth]{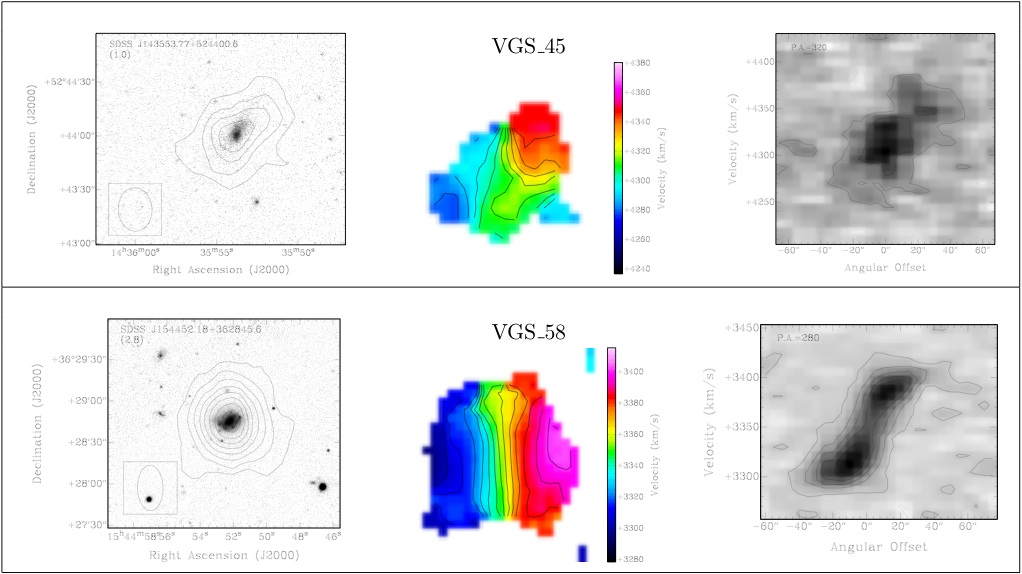}
\caption[]{Targeted void galaxies. Contours in the total intensity maps (left) are at $5 \times 10^{19} cm^{-2}$ plus increments of $10^{20} cm^{-2}$. Confidence level ($\sigma$) of the lowest contour is given in the top left corner of each image. Lines in the velocity field images (center) indicate increments of 8 km s$^{-1}$. Position-Velocity diagrams (right) are along the kinematic major axis, contours are at increments of -1.5 (grey), 1.5 (black) + increments of 3$\sigma$ .
\label{fig:vgs}}
\end{figure}

\begin{figure}
\centering
\includegraphics[scale=1]{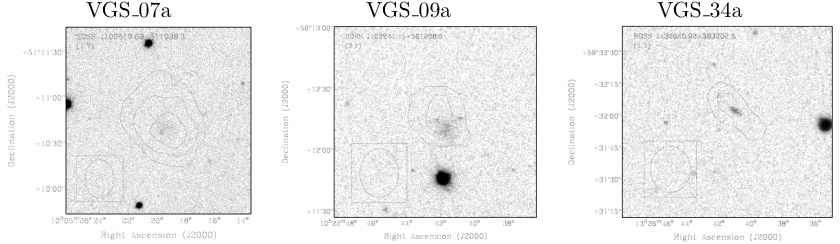}
\caption{Companions.  Note that VGS\_30a, VGS\_38a and VGS\_38b are shown in Figure \ref{fig:vgs} along with the targeted void galaxy. Contours in the total intensity maps are at $5 \times 10^{19} cm^{-2}$ plus increments of $10^{20} cm^{-2}$. Confidence level ($\sigma$) of the lowest contour is given in the top left corner of each image. 
\label{fig:comps}}
\end{figure}

\clearpage

\clearpage

\begin{deluxetable}{ l l l l r r r r r}
\tabletypesize{\scriptsize}
\tablecaption{Parameters of selected void galaxies taken from the SDSS catalog
\label{tab:params}}
\tablewidth{0pt}
\tablehead{
\colhead{Name} & \colhead{SDSS ID} & \colhead{ra} & \colhead{dec} & 
\colhead{$z$} &  \colhead{$g$} & \colhead{$r$} & \colhead{$g-r$} & \colhead{M$_r$} \\
\colhead{} & \colhead{} & \colhead{(J2000.0)} & \colhead{(J2000.0)} & \colhead{} & 
\colhead{} & \colhead{} & \colhead{} 
}
\startdata
VGS\_01 & SDSS J083707.48+323340.8 &  08 37 07.5  & +32 33 41 &   0.018531 & 17.67 & 17.17 & 0.49 & -17.4 \\
VGS\_07 & SDSS J100642.44+511623.9 &  10 06 42.5  & +51 16 24 &   0.016261 & 17.36 & 17.30 & 0.06 & -16.9 \\
VGS\_09 & SDSS J102250.68+561932.1 &  10 22 50.7  & +56 19 32 &   0.012995 & 17.78 & 17.52 & 0.26 & -16.2 \\
VGS\_12 & SDSS J102819.23+623502.6 &  10 28 19.2  & +62 35 03 &   0.017804 & 17.67 & 17.41 & 0.26 & -17.0 \\
VGS\_14 & SDSS J103506.46+550847.5 &  10 35 06.5  & +55 08 48 &   0.013154 & 16.99 & 16.72 & 0.28 & -17.1 \\
VGS\_30 & SDSS J130526.08+544551.9 &  13 05 26.1  & +54 45 52 &   0.019435 & 18.27 & 18.05 & 0.22 & -16.6 \\
VGS\_32 & SDSS J132232.48+544905.5 &  13 22 32.5  & +54 49 06 &   0.011835 & 14.70 & 14.17 & 0.53 & -19.4 \\
VGS\_34 & SDSS J132718.56+593010.2 &  13 27 18.6  & +59 30 10 &   0.016539 & 16.09 & 15.22 & 0.87 & -19.1 \\
VGS\_35 & SDSS J135113.62+453509.2 &  13 51 13.6  & +45 35 09 &   0.017299 & 16.77 & 16.36 & 0.41 & -18.0 \\
VGS\_36 & SDSS J135535.46+593041.3 &  13 55 35.5  & +59 30 41 &   0.022398 & 16.81 & 16.46 & 0.36 & -18.5 \\
VGS\_38 & SDSS J140034.49+551515.1 &  14 00 34.5  & +55 15 15 &   0.013820 & 17.30 & 16.95 & 0.35 & -16.9 \\
VGS\_42 & SDSS J142416.41+523208.3 &  14 24 16.4  & +52 32 08 &   0.018762 & 16.49 & 15.88 & 0.61 & -18.7 \\
VGS\_44 & SDSS J143052.33+551440.0 &  14 30 52.3  & +55 14 40 &   0.017656 & 15.38 & 14.92 & 0.46 & -19.5 \\
VGS\_45 & SDSS J143553.77+524400.6 &  14 35 53.8  & +52 44 01 &   0.014553 & 17.60 & 17.33 & 0.26 & -16.7 \\
VGS\_58 & SDSS J154452.18+362845.6 &  15 44 52.2  & +36 28 46 &   0.011522 & 16.05 & 15.70 & 0.35 & -17.8 \\
\enddata
\tablecomments{Units of right ascension are hours, minutes, and 
seconds, and units of declination are degrees, arcminutes, and arcseconds. 
 $z$ is the spectroscopic redshift.  $g$ and $r$ list the apparent model magnitudes available in the SDSS DR7 catalog.  Absolute magnitudes have been corrected for galactic extinction.  }
\end{deluxetable}

\begin{deluxetable}{ r c c c c c c c c }
\tabletypesize{\scriptsize}
\tablecaption{Parameters of the voids surrounding the target void galaxies.  \label{tab:voidparams}}
\tablewidth{0pt}
\tablehead{
\colhead{Name} & \colhead{R$_{void}$} & \colhead{D$_{gal,void}$} & \colhead{DR} &
\colhead{$\delta$} & \colhead{Nearest} & \colhead{Nearest} &
\colhead{Void Name} \\
\colhead{} & \colhead{(Mpc)} & \colhead{(Mpc)} & \colhead{} & \colhead{} & 
\colhead{nbr} & \colhead{nbr6} & \colhead{} \\
\colhead{(1)} & \colhead{(2)} & \colhead{(3)} & \colhead{(4)} & 
\colhead{(5)} & \colhead{(6)} & \colhead{(7)} & \colhead{(8)} & 
}
\startdata
VGS\_01 & 16.72 & 18.48 & 1.11 & -0.68 & 1.92 & 2.77 & UrsaMajor \\
VGS\_07 & 16.72 & 13.39 & 0.80 & -0.82 & 3.97 & 4.67 & UrsaMajor \\
VGS\_09 & 16.50 & 19.77 & 1.20 & -0.67 & 1.52(4) & 4.09 & Leo(Top) \\
VGS\_12 & 18.64/16.72 & 17.22/17.68 & 0.92/1.06 & -0.78 & 3.51 & 4.01 & UrsaMinorII/UrsaMajor \\
VGS\_14 & 16.50 & 18.37 & 1.11 & -0.68 & 1.52(2) & 4.18 & Leo(Top) \\
VGS\_30 & 16.25 & 12.69 & 0.78 & -0.85 & 3.31 & 5.16 & UrsaMinorII \\
VGS\_32 & 18.64 & 17.26 & 0.93 & -0.74 & 3.69 & 4.65 & UrsaMinorI \\
VGS\_34 & 18.64 & 9.65 & 0.52 & -0.66 & 3.39 & 4.19 & UrsaMinorI \\
VGS\_35 & 16.25 & 12.71 & 0.78 & -0.57  & 2.82 & 4.07 & UrsaMinorII \\
VGS\_36 & 16.25/18.64 & 15.66/22.03 & 0.96/1.18 & -0.74 & 2.86 & 3.67 & UrsaMinorII(Top)/UrsaMinorI \\
VGS\_38 & 18.64 & 14.98 & 0.80 & -0.58 & 1.61 & 3.50 & UrsaMinorI \\
VGS\_42 & 18.64/16.25 & 19.25/12.79 & 1.03/0.79 & -0.66 & 1.87 & 3.80 & UrsaMinorI/UrsaMinorII \\
VGS\_44 & 18.64 & 15.70 & 0.52 & -0.83 & 2.99 & 4.13 & UrsaMinorI \\
VGS\_45 & 18.64 & 16.50 & 0.88 & -0.38 & 3.10 & 3.44 & UrsaMinorI \\
VGS\_58 & 14.57 & 12.59 & 0.86 & -0.88 & 0.34 & 5.26 & CoronaBorealisII \\
\enddata
\tablecomments{   
~Listed are:  name of the target void galaxy (1), 
(equivalent) void radius $R_{void}$ (2), the distance $D_{gal,void}$ of the galaxy from the void center (3), ratio of 
void center distance to void radius (4), the filtered density contrast $\delta$ at R$_f=1$h$^{-1}$Mpc (5), 
the distance to the nearest neighbor (6), the average distance of the six nearest neighbors (7) and 
the void name taken from \cite{Fairall1998} and \cite{Hoyle2002a} (8). See Section \ref{sec:sample} for further specification.}
\end{deluxetable}

\begin{table}
\begin{center}
\tabletypesize{\scriptsize}
\caption{Parameters of the WSRT observations
\label{tab:obsparams}}
\begin{tabular}{l l}
\tableline
\tableline
Configuration 		& 	Maxi-short \\
Date 			& 	2006,2007 \\
No. telescopes 		& 	13 \\
Exposure time 		& 	12 h \\
Total bandwidth 	& 	40 MHz \\
No. channels 		& 	512 \\
Shortest spacing 	& 	36 m\\
Longest spacing 	& 	2754 m\\
FWHM primary beam & 	36$^{\prime}$ \\
Synthesized beam 	&  	19$^{\prime\prime} \times$ 19$^{\prime\prime}$/sin($\delta$) \\
rms 		 		& 	0.4 mJy Beam$^{-1}$ \\
velocity resolution 	&	8.6 km s$^{-1}$ \\
\tableline
\end{tabular}
\end{center}
\end{table}

\begin{deluxetable}{ l l l l r r r r c c c  }
\tabletypesize{\scriptsize}
\tablecaption{Companion galaxies parameters taken from the SDSS catalog
\label{tab:company}}
\tablewidth{0pt}
\tablehead{
\colhead{name} & \colhead{SDSS ID} &
\colhead{ra} & \colhead{dec} & 
\colhead{$g$} &\colhead{$r$} & \colhead{$g - r$} &
\colhead{M$_r$} & 
\colhead{$\Delta \theta$} & \colhead{$\Delta$d} &
\colhead{$\Delta$ v}  \\
\colhead{} & \colhead{} & 
\colhead{(J2000.0)} & \colhead{(J2000.0)} & 
\colhead{} & \colhead{} & \colhead{} & \colhead{} &
\colhead{($^\prime$)} & \colhead{(kpc)} &
\colhead{(km s$^{-1}$)} 
}
\startdata
VGS\_07a & SDSS J100519.69+511038.3 &  10 05 19.7  & +51 10 38 & 20.15 &  20.12 &  0.03 &  -14.1 &  14.2 &   288 &          -21 \\
VGS\_09a & SDSS J102241.41+561208.5  &  10 22 41.4  & +56 12 09 & 22.29 &  22.26 &  0.02 &  -11.4 &   7.5 &   121 &          -85 \\
VGS\_30a & SDSS J130531.13+544553.8  &  13 05 31.1  & +54 45 54 & 18.67 &  18.39 &  0.29 &  -16.2 &   0.7 &    17 &          -74 \\
VGS\_34a & SDSS J132640.92+593202.5  &  13 26 40.9  & +59 32 03 & 20.41 &  20.31 &  0.10 &  -14.0 &   5.1 &   104 &           42 \\
VGS\_38a & SDSS J140032.44+551445.9  &  14 00 32.4  & +55 14 46 & 17.74 &  17.59 &  0.15 &  -16.1 &   0.6 &     9 &          -21 \\
VGS\_38b & SDSS J140025.68+551318.5  &  14 00 25.7  & +55 13 19 & 19.02 &  18.82 &  0.20 &  -14.9 &   2.3 &    37 &           10 \\
\enddata
\tablecomments{Units of right ascension are hours, minutes, and 
seconds, and units of declination are degrees, arcminutes, and arcseconds.   $g$ and $r$ list the apparent model magnitudes as measured by the SDSS DR7. Absolute magnitudes have been corrected for galactic extinction.  $\Delta \theta$, $\Delta$d and $\Delta$v list the displacement from the beam center, projected sky separation, and velocity separation, respectively, between the target and companion galaxy.}
\end{deluxetable}

\begin{deluxetable}{ l r r r r r r r r r r }
\tabletypesize{\scriptsize}
\tablecaption{H \textsc{i} properties of targeted void galaxies and companions
\label{tab:vgs}} 
\tablewidth{0pt}
\tablehead{
\colhead{Name} & \colhead{$M_{\textrm{H \textsc{I}}}$} & \colhead{$V_{sys}$} & \colhead{D} & 
\colhead{r$_{90}$} & \colhead{r$_{\textrm{H \textsc{I}}}$} & 
\colhead{$W_{50}$} & \colhead{$W_{20}$} &  \colhead{i} & \colhead{$M_{dyn}$}  & \colhead{$M_{\textrm{H \textsc{I}}}/L_r$}  \\
\colhead{} & \colhead{($10^8 M_\odot$)} & \colhead{(km s$^{-1}$)} & \colhead{(Mpc)} &
\colhead{(kpc)} & \colhead{(kpc)} & \colhead{(km s$^{-1}$)} & \colhead{(km s$^{-1}$)} &  \colhead{($^\circ$)} &  
\colhead{($10^{10} M_\odot$)} & \colhead{} 
}
\startdata

VGS\_01 &    $<$  2.1 \; \; &            - &   - &  2.4 &  - &            - &           - & 60 &  - &   - \\
VGS\_07 &   8.6 $\pm$  0.8 &         4901 &  70 &  2.6 &  9.2 &          119 &          144 &           50 &     1.28 &   1.9 \\
VGS\_07a & 3.23 $\pm$ 0.65 &     4880 &       69 & 3.5 &  -  &           41 &           49 &  -  &  -  &  9.7 \\ 
VGS\_09 &  10.3 $\pm$  1.0 &         3881 &  55 &  3.1 &  9.1 &          110 &          126 &           68 &     0.74 &   4.4 \\
VGS\_09a & 0.57 $\pm$ 0.14 &     3796 &       54 & 1.5 &  -  &           41 &           49 &  -  &  -  & 20.2 \\ 
VGS\_12 &  29.9 $\pm$  4.2 &         5316 &  76 &  2.4 & 18.0 &          154 &          179 &           29 &    10.55 &   6.2 \\
VGS\_14 &   6.4 $\pm$  1.0 &         3933 &  56 &  3.3 &  8.7 &          110 &          134 &           67 &     0.72 &   1.3 \\
VGS\_30 &   5.5 $\pm$  1.0 &         5666 &  81 &  3.7 &  $<$  7.5 &           50 &           93 &           77 &  $<$     0.12 &   1.7 \\
VGS\_30a & 4.52 $\pm$ 0.79 &     5592 &       79 & 2.1 &  -  &           24 &           41 &  -  &  -  &  2.1 \\ 
VGS\_32 &  38.0 $\pm$  4.5 &         3522 &  50 &  4.1 & 19.0 &          171 &          187 &           46 &     6.29 &   0.9 \\
VGS\_34 &  23.9 $\pm$  2.9 &         4917 &  70 &  3.6 & 10.2 &          231 &          299 &           50 &     5.34 &   0.7 \\
VGS\_34a & 0.50 $\pm$ 0.16 &     4959 &       70 & 1.1 &  -  &           33 &           58 &  -  &  -  &  1.7 \\ 
VGS\_35 &  10.7 $\pm$  1.3 &         5191 &  74 &  3.0 & 10.8 &          145 &          187 &           65 &     1.61 &   0.9 \\
VGS\_36 &  19.8 $\pm$  2.7 &         6684 &  95 &  5.0 & 13.0 &          190 &          224 &           79 &     2.82 &   1.1 \\
VGS\_38 &   9.1 $\pm$  0.7 &         3853 &  55 &  2.8 &  6.9 &           50 &           92 &           39 &     0.26 &   2.0 \\
VGS\_38a & 0.86 $\pm$ 0.14 &     3832 &       54 & 1.3 &  -  &           67 &           75 &  -  &  -  &  0.4 \\ 
VGS\_38b & 1.39 $\pm$ 0.22 &     3863 &       55 & 1.4 &  -  &           41 &           66 &  -  &  -  &  2.0 \\ 
VGS\_42 &   4.0 $\pm$  1.5 &         5601 &  80 &  3.5 &  $<$  8.1 &          128 &          171 &           58 &  $<$     1.08 &   0.2 \\
VGS\_44 &   4.9 $\pm$  1.1 &         5295 &  76 &  3.6 &  $<$  7.0 &           76 &          110 &           31 &  $<$     0.90 &   0.1 \\
VGS\_45 &   3.5 $\pm$  1.3 &         4316 &  62 &  4.0 &  $<$  6.9 &           68 &          102 &           66 &  $<$     0.22 &   1.0 \\
VGS\_58 &   7.0 $\pm$  0.6 &         3351 &  48 &  2.6 &  7.4 &          143 &          151 &           38 &     2.39 &   0.7 \\

\enddata
\tablecomments{Non-detections list the 3$\sigma$ upper limit on the H \textsc{i} mass. $V_{sys}$ is the systemic H \textsc{i} velocity, using the optical definition.  D is the distance to the target galaxy.  r$_{90}$ and r$_{\textrm{H \textsc{I}}}$ list the optical and H \textsc{i} radius, respectively.  When possible, poorly resolved systems list the upper limit for r$_{\textrm{H \textsc{I}}}$ and $M_{dyn}$.  $W_{50}$ and $W_{20}$ are the 50\% and 20\% H \textsc{i} line widths, respectively, corrected for instrumental broadening.  The inclination, $i$,  is calculated such that 90$^\circ$ is edge-on.}
\end{deluxetable}

\begin{deluxetable}{l r r r r r r}
\tabletypesize{\scriptsize}
\tablecaption{Stellar and star formation parameters for void galaxies
\label{tab:starmass}}
\tablewidth{0pt}
\tablehead{
\colhead{Name} & \colhead{$M_{*}$} & 
 \colhead{SFR$_{H\alpha}$} &  \colhead{SFR$_{1.4 GHz}$} &
 \colhead{EW(H$\alpha$)}  &
 \colhead{SFR$_{H\alpha}/M_{*}$} & \colhead{SFR$_{H\alpha}$/$M_{\textrm{H \textsc{I}}}$} \\
\colhead{} & 
\colhead{($10^8 M_\odot$)} & 
\colhead{($M_\odot$ yr$^{-1}$)} & \colhead{($M_\odot$ yr$^{-1}$)} & \colhead{(\AA)} &
\colhead{($10^{-11}$ yr$^{-1}$)} & \colhead{($10^{-11}$ yr$^{-1}$)}
}
\startdata
VGS\_01 &    6.8 & 1.43 & $<$ 0.16 &    55.3 &  211.5 &   -  \\
VGS\_07 &    0.4 & 0.25 & 0.20 &   202.9 &  604.4 &  28.6 \\
VGS\_09 &    0.4 & 0.00 & $<$ 0.08 &    16.7 &    0.0 &   0.0 \\
VGS\_12 &   10.5 & 0.05 & $<$ 0.15 &    22.4 &    5.2 &   1.8 \\
VGS\_14 &    2.0 & 0.14 & $<$ 0.08 &    53.4 &   70.7 &  21.7 \\
VGS\_30 &    0.7 & 0.17 & $<$ 0.17 &    11.3 &  241.8 &  31.5 \\
VGS\_32 &   68.6 & - & 0.57 &    15.1 &    0.0 &   0.0 \\
VGS\_34 &   75.4 & 0.95 & 2.97 &    32.2 &   12.6 &  39.7 \\
VGS\_35 &   12.8 & 0.07 & $<$ 0.14 &    19.7 &    5.6 &   6.7 \\
VGS\_36 &   19.4 & 0.17 & $<$ 0.23 &    17.0 &    8.7 &   8.5 \\
VGS\_38 &    0.5 & 0.00 & $<$ 0.08 &    -0.3 &    0.0 &   0.0 \\
VGS\_42 &   38.9 & 0.11 & $<$ 0.16 &     9.5 &    2.9 &  28.3 \\
VGS\_44 &   21.1 & 1.47 & 0.95 &    85.9 &   69.8 & 302.4 \\
VGS\_45 &    0.6 & 0.00 & $<$ 0.10 &    11.4 &    0.0 &   0.0 \\
VGS\_58 &   10.0 & 0.11 & $<$ 0.06 &    10.9 &   11.3 &  16.1 \\
\enddata
\tablecomments{Targets not detected in the 1.4 GHz continuum list the  3$\sigma$ upper limit on the SFR$_{1.4 GHz}$.}
\end{deluxetable}

\begin{deluxetable}{r l l l r r r r r r r}
\tabletypesize{\scriptsize}
\tablecaption{Control sample galaxy parameters taken from the SDSS catalog
\label{tab:altifs1}}
\tablewidth{0pt}
\tablehead{
\colhead{\#} & \colhead{SDSS ID} & \colhead{ra} & \colhead{dec} & 
\colhead{$z$} &  \colhead{$g$} & \colhead{$r$} & \colhead{$g-r$} & \colhead{M$_r$} & 
\colhead{$\delta$} & \colhead{$\Delta \theta$} \\
\colhead{} & \colhead{} & \colhead{(J2000.0)} & \colhead{(J2000.0)} & \colhead{} & 
\colhead{} & \colhead{} & \colhead{} & \colhead{} & \colhead{} & \colhead{($^{\prime}$)} 
}
\startdata
1 & SDSS J083649.77+325109.3  &   08 36 49.8  & +32 51 09
 &   0.025672 &  16.26 &  15.90 &  0.36 &  -19.4 &  0.28 &  17.9  \\
2 & SDSS J102002.03+562423.8  &   10 20 02.0  & +56 24 24
 &   0.024881 &  17.53 &  17.22 &  0.31 &  -18.0 &  7.81 &  23.8  \\
3 & SDSS J102257.38+555449.3  &   10 22 57.4  & +55 54 50
 &   0.025082 &  15.63 &  15.08 &  0.55 &  -20.1 &  7.81 &  24.7  \\
4 & SDSS J102316.27+562656.2  &   10 23 16.3  & +56 26 56
 &  -   &  18.57 &  18.50 &  0.07 &  -16.6 &  6.84 &   8.2  \\
5 & SDSS J102653.21+622011.4  &   10 26 53.2  & +62 20 12
 &   0.031975 &  15.41 &  14.78 &  0.64 &  -20.9 &  -0.44 &  17.9  \\
6 & SDSS J102823.88+623706.6  &   10 28 23.9  & +62 37 07
 &   0.030462 &  16.39 &  15.60 &  0.78 &  -20.0 &  1.84 &   2.2  \\
7 & SDSS J102942.07+624826.2  &   10 29 42.1  & +62 48 27
 &   0.023326 &  16.64 &  16.24 &  0.40 &  -18.8 &  1.63 &  16.5  \\
8 & SDSS J103105.76+623531.5  &   10 31 05.8  & +62 35 32
 &   0.030938 &  18.09 &  17.73 &  0.36 &  -17.9 &  3.65 &  19.2  \\
9 & SDSS J130418.89+544209.9  &   13 04 18.9  & +54 42 10
 &    - &  18.81 &  18.50 &  0.30 &  -15.7 &  3.23 &  10.4  \\
10 & SDSS J132433.97+592018.6  &   13 24 34.0  & +59 20 19
 &   0.028574 &  15.29 &  14.38 &  0.91 &  -21.1 &  0.04 &  23.1  \\
11 & SDSS J132523.37+593643.1  &   13 25 23.4  & +59 36 43
 &   0.026625 &  15.10 &  14.65 &  0.46 &  -20.7 &  -0.13 &  16.0  \\
12 & SDSS J132630.72+594313.7 &   13 26 30.7  & +59 43 14
 &   0.029162 &  15.30 &  14.48 &  0.83 &  -21.1 &  0.21 &  14.4  \\
13 & SDSS J132722.97+592537.8  &   13 27 23.0  & +59 25 38
 &   0.028064 &  16.96 &  16.80 &  0.16 &  -18.7 &  0.31 &   4.6  \\
14 & SDSS J135839.08+552910.3  &   13 58 39.1  & +55 29 10
 &   0.027418 &  18.06 &  17.86 &  0.20 &  -17.5 &  -0.70 &  21.5  \\
15 & NGC 5422 &   14 00 42.0  & +55 09 52
 &   0.006071 &  12.93 &  12.06 &  0.87 &  -20.1 & 24.71 &   5.5  \\
16 & SDSS J140058.24+553405.1  &   14 00 58.2  & +55 34 05
 &   0.006174 &  16.37 &  15.90 &  0.47 &  -16.2 & 24.71 &  19.1  \\
17 & SDSS J140101.95+545555.2  &   14 01 02.0  & +54 55 55
 &   0.005950 &  17.31 &  16.69 &  0.61 &  -15.4 & 17.97 &  19.7  \\
18 & SDSS J142515.68+522128.9  &   14 25 15.7  & +52 21 29
 &   0.031980 &  17.84 &  17.56 &  0.29 &  -18.2 &  -0.32 &  14.0  \\
19 & SDSS J143422.81+522839.0  &   14 34 22.8  & +52 28 39
 &   - &  16.51 &  16.26 &  0.25 &  -17.2 &  1.48 &  20.6  \\
20 & SDSS J143441.11+523935.1  &   14 34 41.1  & +52 39 35
 &   - &  18.32 &  18.14 &  0.18 &  -15.3 &  1.48 &  11.8  \\
21 & SDSS J143519.21+522745.0  &   14 35 19.2  & +52 27 45
 &   - &  18.32 &  18.02 &  0.29 &  -15.4 &  1.48 &  17.0  \\
22 & SDSS J143744.16+524331.2  &   14 37 44.2  & +52 43 32
 &   0.011288 &  15.87 &  15.69 &  0.18 &  -17.8 &  1.48 &  16.8  \\
23  & SDSS J143835.24+524223.1 &   14 38 35.2  & +52 42 23
 &   0.011301 &  15.63 &  15.15 &  0.48 &  -18.3 &  1.48 &  24.5  \\
\enddata
\tablecomments{To be compared with Table \ref{tab:params}. Units of right ascension are hours, minutes, and 
seconds, and units of declination are degrees, arcminutes, and arcseconds. $z$ is the spectroscopic redshift.  
 $g$ and $r$ list the apparent magnitudes as measured by the SDSS DR7.
Absolute magnitudes have been corrected for galactic extinction.  
$\delta$ indicates the filtered density contrast, described in Section \ref{sec:sample}. $\Delta \theta$ lists the angular displacement from the beam center. Control galaxies \# 4, 9, 19, 20 and 21 are not spectroscopically targeted by the SDSS.}
 \end{deluxetable}
 
 \begin{deluxetable}{ r c c r r r r r r r r}
\tabletypesize{\scriptsize}
\tablecaption{H \textsc{i} properties of control sample galaxies
\label{tab:altifs2}} 
\tablewidth{0pt}
\tablehead{
\colhead{\#} & \colhead{$M_{\textrm{H \textsc{I}}}$} & \colhead{$V_{sys}$} & \colhead{D} & 
\colhead{r$_{90}$} & \colhead{r$_{\textrm{H \textsc{I}}}$} & 
\colhead{$W_{50}$} & \colhead{$W_{20}$} &   \colhead{i} & \colhead{$M_{dyn}$}  & \colhead{$M_{\textrm{H \textsc{I}}}/L_r$}  \\
\colhead{} & \colhead{($10^8 M_\odot$)} & \colhead{(km s$^{-1}$)} & \colhead{(Mpc)} &
\colhead{(kpc)} & \colhead{(kpc)} & \colhead{(km s$^{-1}$)} & \colhead{(km s$^{-1}$)} &  \colhead{($^\circ$)} &  
\colhead{($10^{10} M_\odot$)} & \colhead{} 
}
\startdata

1  &  43.0 $\pm$  7.7 &         7702 &          110 &  6.4 & 19.7 &          194 &          229 &           70 &     4.89 &  1.0 \\
2  &  18.6 $\pm$ 10.8 &         7464 &          107 &  5.8 &  $<$  9.8 &           59 &          119 &           26 &     $<$ 1.05 &  1.6 \\
3  &  24.2 $\pm$ 16.3 &         7525 &          107 & 10.1 &  -  &          213 &          260 &           61 &     - &  0.3 \\
4  &   5.8 $\pm$  3.4 &         7350 &          105 &  3.6 &  -  &           39 &           51 &           74 &     - &  1.7 \\
5 & 137.6 $\pm$ 79.0 &         9593 &          137 &  8.5 & 37.9 &          187 &          148 &           30 &    30.02 &  0.8 \\
6  &   $<$  7.3 &         - &          131 &  3.9 &  - &            - &            - &           70 &     - & - \\
7 &  20.4 $\pm$  7.6 &         6998 &          100 &  3.2 & 15.0 &           50 &          202 &           46 &     0.43 &  0.8 \\
8  &   $<$ 18.1 &         - &          133 &  5.2 &  - &            - &            - &           78 &     - & -  \\
9  &   2.3 $\pm$  1.2 &         4765 &           68 &  2.4 &  $<$  4.3 &          105 &          114 &           72 &     $<$ 0.31 &  1.6 \\
10  &   $<$ 24.0 &         - &          122 &  9.5 &  - &            - &            - &           68 &     - & -  \\
11  & 105.0 $\pm$ 21.6 &         7987 &          114 &  6.3 & 22.7 &           73 &          150 &           43 &     1.50 &  0.8 \\
12 &   $<$ 10.8 &         - &          125 &  9.4 &  - &            - &            - &           58 &     - & -  \\
13  &  36.5 $\pm$  6.3 &         8419 &          120 &  6.8 & 15.7 &           95 &          160 &           50 &     1.42 &  1.7 \\
14 &  13.3 $\pm$  9.0 &         8225 &          118 &  4.3 &  -    &          134 &          155 &           63 &     - &  1.8 \\
15 &   0.6 $\pm$  0.9 &         1821 &           26 &  3.8 &  -    &            - &            - &           57 &     - &  0.0 \\
16  &   $<$  0.7 &         - &           26 &  1.9 &  - &            - &            - &           60 &     - & -  \\
17 &   $<$  0.7 &         - &           25 &  1.1 &  - &            - &            - &           25 &     - & -  \\
18  &  35.0 $\pm$ 19.7 &         9594 &          137 &  2.9 & 20.6 &          206 &          214 &           65 &     6.21 &  2.6 \\
19 &  21.1 $\pm$  6.8 &         3386 &           48 &  3.8 & 11.5 &           96 &          117 &           30 &     2.54 &  3.8 \\
20 &   3.2 $\pm$  1.5 &         3386 &           48 &  1.5 &  -  &           96 &          108 &           63 &     - &  3.2 \\
21 &   2.5 $\pm$  0.8 &         3386 &           48 &  2.6 &  -  &           58 &           71 &           60 &     - &  2.2 \\
22  &  15.6 $\pm$  5.6 &         3386 &           48 &  3.0 & 10.1 &           54 &           87 &           51 &     0.29 &  1.6 \\
23 &  $<$  4.5 &         - &           48 &  3.0 &  - &            - &            - &           66 &     - & -  \\

\enddata
\tablecomments{To be compared with Table \ref{tab:vgs}. Non-detections list the 3$\sigma$ upper limit on the H \textsc{i} mass. 
$V_{sys}$ is the systemic H \textsc{i} velocity, using the optical definition.  D is the distance to the target galaxy.  r$_{90}$ and r$_{\textrm{H \textsc{I}}}$ list the optical and H \textsc{i} radius, respectively.  When possible, poorly resolved systems list the upper limit for r$_{\textrm{H \textsc{I}}}$ and $M_{dyn}$.  $W_{50}$ and $W_{20}$ are the 50\% and 20\% H~\textsc{i} line widths, respectively, corrected for instrumental broadening.  The inclination, $i$, is calculated such that 90$^\circ$ is edge-on.}
 \end{deluxetable}

\begin{deluxetable}{r r r r r r r }
\tabletypesize{\scriptsize}
\tablecaption{Stellar and star formation parameters for control sample galaxies
\label{tab:starmass2}}
\tablewidth{0pt}
\tablehead{
\colhead{\#} & \colhead{$M_{*}$} &
 \colhead{SFR$_{H\alpha}$} &
 \colhead{SFR$_{1.4GHz}$} &
 \colhead{EW(H$\alpha$)}  &
 \colhead{SFR$_{H\alpha}/M_{*}$} & \colhead{SFR$_{H\alpha}$/$M_{\textrm{H \textsc{I}}}$} \\
\colhead{} & 
\colhead{($10^8 M_\odot$)} &
\colhead{($M_\odot$ yr$^{-1}$)} & \colhead{($M_\odot$ yr$^{-1}$)} &
\colhead{(\AA)}&
\colhead{($10^{-11}$ yr$^{-1}$)} & \colhead{($10^{-11}$ yr$^{-1}$)}
}
\startdata
1 &   55.0 & 0.36 & $<$ 0.67 &    16.9 &    6.6 &   8.4 \\
2 &    8.5 & 0.28 & $<$ 1.19 &    35.1 &   32.4 &  14.8 \\
3  &  302.0 & 0.21 & $<$ 1.37 &     2.6 &    0.7 &   8.8 \\
4   &    - & - & $<$ 0.33 &     - &    - &   - \\
5  &  407.4 & 0.70 & 1.32 &     3.2 &    1.7 &   5.1 \\
6  &  354.8 & 0.00 & $<$ 0.85 &    -1.2 &    0.0 &  - \\
7  &   20.0 & 0.27 & $<$ 0.49 &    24.8 &   13.8 &  13.5 \\
8  &    - & - & $<$ 2.10 &    12.4 &    - &  - \\ 
9  &    - & - & $<$ 0.15 &     - &    - &   - \\
10  &    - & - & $<$ 2.78 &     0.1 &    - &  -  \\
11  &    - & - & 12.29 &     - &    - &   - \\
12 &  691.8 & 7.14 & 9.33 &    46.2 &   10.3 &   - \\
13  &    4.5 & 0.38 & $<$ 0.39 &    44.1 &   84.6 &  10.3 \\
14  &    - & - & $<$ 1.10 &    40.1 &    - &   - \\
15 &    - & - & $<$ 0.02 &     - &    - &   - \\
16  &    3.2 & 0.00 & $<$ 0.08 &     2.4 &    0.0 &  -  \\
17  &    2.8 & 0.00 & $<$ 0.08 &    -1.1 &    0.0 &  -  \\
18  &   20.0 & 0.70 & $<$ 0.76 &   134.7 &   34.9 &  19.9 \\
19  &    - & - & $<$ 0.17 &     - &    - &   - \\
20  &    - & - & $<$ 0.08 &     - &    - &   - \\
21 &    - & - & $<$ 0.12 &     - &    - &   - \\
22  &    0.2 & 0.10 & $<$ 0.12 &   120.5 &  641.4 &   6.5 \\
23 &    - & - & $<$ 0.52 &     5.2 &    - &  -  \\
\enddata
\tablecomments{To be compared with Table \ref{tab:starmass}. Targets not detected in the 1.4 GHz continuum list the  3$\sigma$ upper limit on the SFR$_{1.4 GHz}$.}
\end{deluxetable}

\begin{deluxetable}{l c c c}
\tabletypesize{\scriptsize}
\tablecaption{Stellar and star formation properties as compared with \cite{Rojas2005}
\label{tab:rojas}}
\tablewidth{0pt}
\tablehead{
\colhead{Property} & \colhead{Our void sample} & \colhead{Rojas void sample} & \colhead{Rojas wall sample} \\
\colhead{} & \colhead{($\mu \pm \sigma_\mu$)} & \colhead{($\mu \pm \sigma_\mu$)} & \colhead{($\mu \pm \sigma_\mu$)}
}
\startdata
EW(H$\alpha$)(\AA) 				& 32.143 $\pm$ 7.90   		& 33.316 $\pm$ 3.74 	& 21.91 $\pm$ 0.809 \\
log$_{10}$($M_*$/$M_\odot$) 		& 9.217 $\pm$ 0.152	 	& 9.333 $\pm$ 0.066 	& 9.390  $\pm$  0.018 \\
SFR(H$\alpha$)($M_\odot$ yr$^{-1}$) 	& 0.451 $\pm$ 0.188		& 0.323  $\pm$ 0.048	& 0.194  $\pm$ 0.011 \\
S-SFR(H$\alpha$)(yr$^{-1}$) 			& (39.81 $\pm$ 20.9) $\times 10^{-11}$ & (17.57 $\pm$ 3.222) $\times 10^{-11}$ & (12.57 $\pm$ 0.958) $\times 10^{-11}$ \\
\enddata
\end{deluxetable}

\end{document}